\documentclass[prd,floatfix,onecolumn,amsmath,amssymb,floatfix]{revtex4}
\usepackage{graphicx,color,dcolumn,booktabs,bm}
\usepackage{subfigure}
\usepackage{longtable,lscape}
\usepackage{amssymb}
\usepackage{indentfirst}
\usepackage{epsfig}
\usepackage{feynmf}   %{feynmp}
\usepackage{slashed}  %for Feynman symbols
\usepackage{cases}
\definecolor{maroon}{RGB}{139,25,150}%burada 0-255 arasi her biri icin numara vererek renk elde et
\usepackage{multirow}
\usepackage{float}
\usepackage{graphicx,color,dcolumn,booktabs,bm}
\usepackage[colorlinks, citecolor=blue,anchorcolor=red,menucolor=red, linkcolor=red,filecolor=red,runcolor=red,urlcolor=blue,frenchlinks=red, urlcolor=blue]{hyperref}
\usepackage{orcidlink}

\allowdisplaybreaks

\begin{document}
%%%%%%%%%%%%%%%%%%%%%%%%%%%%%%%%%

\preprint{}
\preprint{}
\title{\color{blue}{
		
		Semileptonic decays of doubly charmed or bottom baryons to single heavy baryons}}

\author{M.~Shekari Tousi$^{a}$\orcidlink{0009-0007-7195-0838}}

	\author{K.~Azizi$^{a,b}$\orcidlink{0000-0003-3741-2167}} 
	\email{kazem.azizi@ut.ac.ir} \thanks{Corresponding author}

\affiliation{
		$^{a}$Department of Physics, University of Tehran, North Karegar Avenue, Tehran 14395-547, Iran\\
		$^{b}$Department of Physics,  Dogus University,  Dudullu-\"{U}mraniye, 34775 Istanbul, T\"urkiye}

\date{\today}

\begin{abstract}

We investigate the semileptonic decays of baryons containing double charm or double bottom quarks, focusing on their transitions to single heavy baryons through three-point QCD sum rule framework. In our calculations, we take into account nonperturbative operators with mass dimensions up to five. We calculate the form factors associated with these decays, emphasizing the vector and axial-vector transition currents in the corresponding amplitude. By applying fitting functions for the form factors based on the squared momentum transfer, we derive predictions for decay widths and branching ratios  in their possible lepton channels. These findings offer valuable insights for experimentalists exploring semileptonic decays of doubly charmed or bottom baryons.  Perhaps they can be validated in upcoming experiments like LHCb. These investigations contribute  to a deeper understanding of the decay mechanisms in these baryonic channels.

\end{abstract}

\maketitle
\section{Introduction}
Since the quark model was introduced \cite{GellMann:1964nj}, efforts have been made to create a comprehensive hadron spectrum that includes all the particles predicted by the model. Recent studies focusing on hadrons with heavy quarks have attracted considerable interest. While the quark model has successfully explained many aspects of hadron spectra, it has yet to confirm all theoretically predicted particles, particularly the doubly heavy baryons. The first such baryon, the $\Xi^+_{cc}(3520)$, was identified by the SELEX collaboration in 2002 through its decay channel \( p D^+ K^- \) \cite{SELEX:2002wqn}, with a subsequent confirmation in 2005 \cite{SELEX:2004lln}. In a landmark achievement, the LHCb experiments validated the existence of the doubly charmed baryon \( \Xi^{++}_{cc} (3621) \) in 2017, utilizing the decay channel \( \Xi^{++}_{cc} \rightarrow \Lambda^+_{c} K^- \pi^+ \pi^+ \) \cite{LHCb:2017iph}, which was later confirmed in 2018 via \( \Xi^{++}_{cc} \rightarrow \Xi^+_{c} \pi^+ \) \cite{LHCb:2018pcs}. In 2019, searches for the \( \Xi^+_{cc} \) were conducted by LHCb through the decay \( \Xi^{+}_{cc} \rightarrow \Lambda^+_{c} K^- \pi^+ \) \cite{LHCb:2019gqy}, and results from this investigation were integrated with findings from 2021 related to the decay \( \Xi^{+}_{cc} \rightarrow \Xi^{+}_{c} \pi^- \pi^+ \), achieving a maximum local significance of 4.0 standard deviations around 3620 MeV for the \( \Xi^{+}_{cc} \) mass \cite{LHCb:2021eaf}. However, this does not signify a substantial gap between the masses of \( \Xi^{+}_{cc} \) and \( \Xi^{++}_{cc} \). Ongoing experimental efforts aim to uncover additional samples of these baryons, yet no new doubly heavy baryons have been discovered thus far \cite{LHCb:2021xba}. Many scientists have engaged in extensive research to specify various properties of these type of baryons, including the masses and residues \cite{ShekariTousi:2024mso, Ebert:2002ig,Zhang:2008rt,Wang:2010hs,Lu:2017meb,Rahmani:2020pol,Yao:2018ifh,Aliyev:2022rrf,Aliev:2012iv,Aliev:2019lvd,Aliev:2012ru,Padmanath:2019ybu,Brown:2014ena,Giannuzzi:2009gh,Shah:2017liu,Shah:2016vmd,Yoshida:2015tia,Valcarce:2008dr,Wang:2010it,Ortiz-Pacheco:2023kjn,Wang:2018lhz},  chiral effective Lagrangians \cite{Qiu:2020omj}, strong coupling constants \cite{Olamaei:2021hjd,Aliev:2021hqq,Aliev:2020lly,Alrebdi:2020rev,Rostami:2020euc,Olamaei:2020bvw,Aliev:2020aon}, strong interactions and decays \cite{Azizi:2020zin,Qin:2021dqo,Xiao:2017dly}, radiative decay \cite{Aliev:2021hqq,Xiao:2017udy,Lu:2017meb,Rahmani:2020pol,Li:2017pxa,Ortiz-Pacheco:2023kjn}, weak decay processes \cite{Tousi:2024usi,Gerasimov:2019jwp,Wang:2017mqp,Zhao:2018mrg,Xing:2018lre,Jiang:2018oak,Gutsche:2019wgu,Gutsche:2019iac,Ke:2019lcf,Cheng:2020wmk,Hu:2020mxk,Li:2020qrh,Han:2021gkl,Wang:2017azm,Shi:2017dto,Zhang:2018llc,Ivanov:2020xmw,Shi:2020qde,Hu:2017dzi,Li:2018epz,Shi:2019hbf,Shi:2019fph,Sharma:2017txj,Patel:2024mfn,Gutsche:2017hux,Gutsche:2018msz}, magnetic moments \cite{Ozdem:2018uue,Ozdem:2019zis}, lifetimes \cite{Berezhnoy:2018bde}, mixing angles \cite{Aliev:2012nn}, etc., employing diverse methodologies. To accurately compute these characteristics, nonperturbative techniques such as QCD sum rules, introduced by Shifman, Vainshtein, and Zakharov in 1979 \cite{Shifman:1978bx, Shifman:1978by}, are essential. This approach is based on the QCD Lagrangian utilizing correlation function through various interpolating currents. This method has made numerous successful predictions regarding hadronic parameters, validated by various experiments, and is considered highly effective \cite{Aliev:2010uy,Aliev:2009jt,Aliev:2012ru,Agaev:2016dev,Azizi:2016dhy,Wang:2007ys}.

In this research, we discuss weak decays of the doubly heavy spin-1/2 baryons $ \Xi^{++}_{cc}$, $ \Xi^{+}_{cc}$, $ \Omega^{+}_{cc}$, $ \Xi^{0}_{bb}$,  $ \Xi^{-}_{bb}$ and $\Omega^{-}_{bb}$, specifically focusing on their transitions to spin-1/2 single heavy baryons. At the quark level, these decays occur through $ c\rightarrow d/s $ or $ b \rightarrow u/c $  transitions, with the two spectator quarks. The $ b \rightarrow c $ transition will not be addressed in this work and will be reserved for future study. We begin by deriving the form factors for these transitions using three-point QCD sum rule method. We then utilize these form factors to estimate the decay widths and branching ratios for semileptonic processes of doubly heavy baryons. Our findings indicate that these decay channels are significant and could be investigated in future experiments such as LHCb.

Previous research has explored transitions through various methodologies, including the QCD sum rule approach with different interpolating currents \cite{Shi:2019hbf} and the light-front framework \cite{Wang:2017mqp},  considering the two leptons, \( e \) and \( \mu \)  the same. In this work, we calculate the pertinent form factors by utilizing a comprehensive set of interpolating currents applicable to both initial and final baryonic states, building upon recently established residues and masses  for doubly heavy baryons \cite{ShekariTousi:2024mso}. Our analysis includes a detailed examination of decay widths and branching ratios across all possible lepton channels. By meticulously optimizing critical parameters, such as the mixing parameter in the interpolating currents, we aim to enhance the precision of our results while minimizing uncertainties.

This paper is organized as follows: In Section \ref{Sec2}, we derive the sum rules that govern the form factors relevant to the semileptonic decay of doubly heavy baryons. Section \ref{Sec3} focuses on the numerical evaluation of these sum rules, showcasing fit functions that reveal how the form factors vary with respect to the squared momentum transfer. In Section \ref{Sec4}, we present our findings on decay widths and branching ratios across all possible lepton channels. Lastly, Section \ref{Sec5} wraps up the discussion, and further calculation details can be found in the Appendix.

\section{EVALUATION OF FORM FACTORS VIA QCD SUM RULE METHOD}~\label{Sec2}
Before delving into the detailed calculations of sum rules to evaluate the weak decay form factors of doubly heavy baryons, we first investigate the ground states of these baryons within the quark model. In the ground state of doubly heavy baryons featuring identical heavy quarks, denoted as $	\Xi_{QQ}^{(*)}$ and $		\Omega_{QQ}^{(*)}$, the two heavy quarks combine to form a diquark with a total spin of 1. When we include the spin-1/2 of the light quark, we can have states with either total spin-1/2 or spin-3/2. Baryons with a star represent the spin-3/2 states, while those without represent the spin-1/2 states. For these configurations, the interpolating current must be symmetric in relation to the exchange of the heavy quark fields. For baryons comprising two different heavy quarks, in addition to the spin-1 diquark case, the diquark can also have a total spin of zero, resulting in overall spin-1/2 states. In these cases, the interpolating currents, typically labeled as $	\Xi_{bc}^{'}$ and $		\Omega_{bc}^{'}$, are antisymmetric with respect to the two heavy quark fields. This work focuses exclusively on spin-1/2 doubly heavy baryons. With those initial points in mind, let’s return to our central issue. We investigate the weak decays of the the doubly heavy baryons considering the following semileptonic decay channels:
\begin{itemize}
	\item The $cc$ sector:
	\begin{align*}
		\Xi_{cc}^{++}(ccu) & \to\Lambda_{c}^{+}(dcu)/\Sigma_{c}^{+}(dcu)/\Xi_{c}^{+}(scu)/\Xi_{c}^{\prime+}(scu),\\
		\Xi_{cc}^{+}(ccd) & \to\Sigma_{c}^{0}(dcd)/\Xi_{c}^{0}(scd)/\Xi_{c}^{\prime0}(scd),\\
		\Omega_{cc}^{+}(ccs) & \to\Xi_{c}^{0}(dcs)/\Xi_{c}^{\prime0}(dcs)/\Omega_{c}^{0}(scs),
	\end{align*}
	\item The $bb$ sector:
	\begin{align*}
		\Xi_{bb}^{0}(bbu) & \to\Sigma_{b}^{+}(ubu),\\
		\Xi_{bb}^{-}(bbd) & \to\Lambda_{b}^{0}(ubd)/\Sigma_{b}^{0}(ubd),\\
		\Omega_{bb}^{-}(bbs) & \to\Xi_{b}^{0}(ubs)/\Xi_{b}^{\prime0}(ubs).
	\end{align*}
\end{itemize}

\begin{figure}[!]
	\includegraphics[width=0.32\columnwidth]{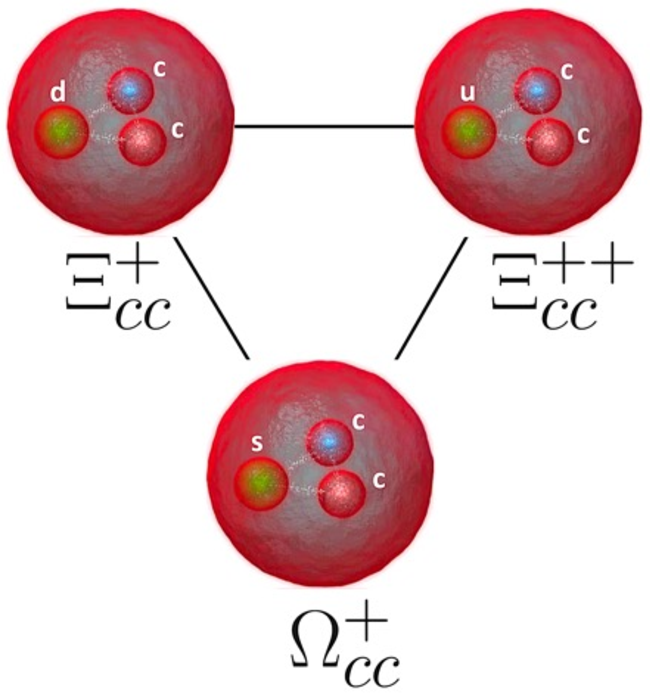}
	\caption{Doubly charm baryons with spin-$1/2$. The case is similar for the doubly bottom baryons and those containing bottom and charm quarks.}
	\label{fig:doubly_heavy_baron}
\end{figure}

\begin{figure}[!]
	\includegraphics[width=0.74\columnwidth]{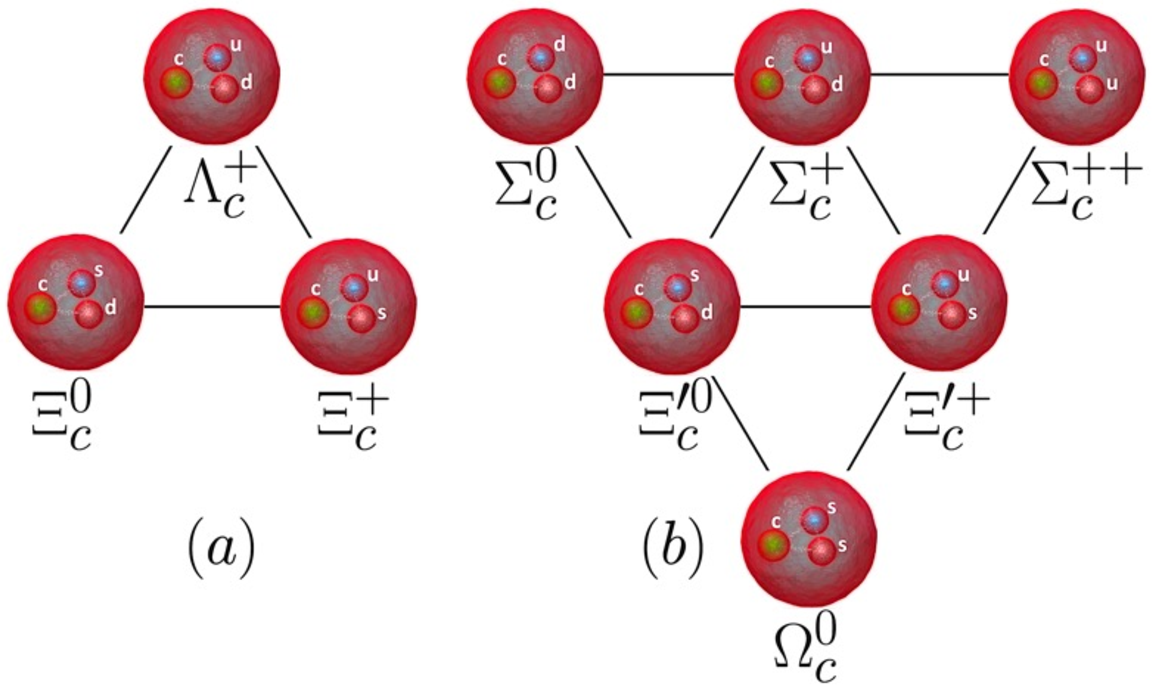}
	\caption{Antitriplets (Panel a) and Sextets (Panel b) of single charm baryons featuring one charm quark and two light quarks. The pattern is similar for the baryons containing a bottom quark. These baryons have a total spin of 1/2, while other sextet exhibiting  spin  3/2  are not shown in the figure.}
	\label{fig:1heavy}
\end{figure}

In these categories, the quark components are clearly indicated in brackets, with the first quarks representing those involved in the decays. Although in our previous study \cite{Tousi:2024usi}, we performed calculations of the decay of $ \Xi^{++}_{cc}\rightarrow \Xi^+_{c} \bar{\ell}\nu_{\ell}$  in two possible lepton channels ( $e^+$ and $\mu ^+$), in this  study we investigate the remaining semileptonic decay channels.
Among the different types of doubly charm baryons, three  can decay via weak interactions: the $	\Xi_{cc}$ isodoublet (ccu, ccd) and the $\Omega_{cc}$ isosinglet (ccs) shown in Fig.~\ref{fig:doubly_heavy_baron}. Similarly, there are three corresponding doubly bottom baryons.
In the decay final state of the  $	\Xi_{cc}$ and $\Omega_{cc}$, we observe baryons that each contain one charm quark. These baryons are organized into antitriplets and sextets of charm baryons, illustrated in Fig.\ref{fig:1heavy}. A similar structure exists for baryons with one bottom quark. The baryons depicted in Fig. \ref{fig:1heavy} possess a total spin of 1/2, while another sextet exhibits a spin of 3/2. This study will concentrate on the 1/2 → 1/2 transition.

To calculate these decay channels, we use the QCD sum rules that utilize a systematic method to evaluate a correlation function via two distinct approaches. The first, known as the physical or phenomenological side, incorporates hadronic degrees of freedom and yields results related to physical quantities like the masses and residues of hadronic states. The second side, known as the QCD or operator product expansion (OPE) side, utilizes the fundamental principles of quantum chromodynamics. This includes aspects such as coupling constants, quark-gluon condensates, and the masses of quarks. By matching conclusions from both sides and concentrating on the coefficients associated with corresponding Lorentz structures, we can establish QCD sum rules for the relevant physical properties.

\subsection{Physical approach}
The considered semileptonic transitions occur via $c\to d/s \, \bar{\ell}\nu_{\ell}$ and $b\to u \, \ell \bar{\nu_{\ell}}$  transitions  (see Fig. \ref{Fig:decay}):

\begin{figure}[htp]
	\includegraphics[width=0.7\columnwidth]{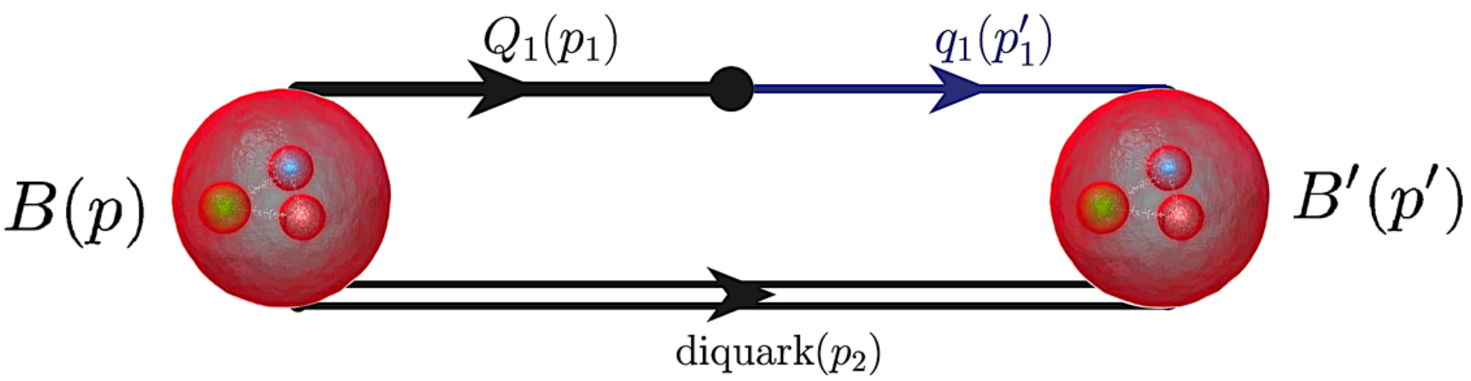}
	 \caption{Feynman diagram depicting the decay of doubly heavy baryons \(B\) into a single heavy baryons \(B^{\prime}\), with two spectator quarks forming a diquark. The momentum of the initial and final baryons are represented as \(p\) and \(p^{\prime}\) respectively. At the quark level, a heavy quark \(Q_{1}\) transits into a light quark \(q_{1}\), and we have two spectator quarks. The black circle symbolizes the weak interaction vertex.}
	\label{Fig:decay}
\end{figure}

To evaluate the transition amplitude, we require the low-energy effective Hamiltonian calculated at the quark level.
The effective Hamiltonian for these semileptonic processes can be written as: 
\begin{eqnarray}\label{heff}
	{\cal H}_{{\rm eff}} & = & \frac{G_{F}}{\sqrt{2}}V_{cs}^{*}[\bar{s}\gamma_{\mu}(1-\gamma_{5})c][\bar{\nu_{\ell}}\gamma^{\mu}(1-\gamma_{5})l],\nonumber \\
	{\cal H}_{{\rm eff}} & = & \frac{G_{F}}{\sqrt{2}}V_{cd}^{*}[\bar{d}\gamma_{\mu}(1-\gamma_{5})c][\bar{\nu_{\ell}}\gamma^{\mu}(1-\gamma_{5})l],\nonumber \\
	\mbox{or}\nonumber \\
	{\cal H}_{{\rm eff}} & = &\frac{G_{F}}{\sqrt{2}}V_{ub}[\bar{u}\gamma_{\mu}(1-\gamma_{5})b][\bar{l}\gamma^{\mu}(1-\gamma_{5})\nu_{\ell}],
\end{eqnarray}
 where the elements \(V_{cs,cd,ub}\) are associated with the Cabibbo-Kobayashi-Maskawa (CKM) matrix and the Fermi coupling constant, referred to as \(G_F\). The decay amplitude is derived by sandwiching  the effective Hamiltonian between the states of the initial and final baryons. These decay processes incorporate both the vector (\(V_\mu\)) and axial vector  (\(A_\mu\)) transitions, each characterized by three form factors. The parameterizations of these transitions are crafted to uphold Lorentz invariance and respect parity constraints, as outlined in \cite{Azizi:2018axf}:

\begin{eqnarray}\label{Cur.with FormFac.}
&&\langle B^{\prime}(p',s')|V^{\mu}| B (p,s)\rangle = \bar
u_{ B^{\prime}}(p',s') \Big[F_1(q^2)\gamma^{\mu}+F_2(q^2)\frac{p^{\mu}}{m_{ B}}
+F_3(q^2)\frac{p'^{\mu}}{m_{ B^{\prime}}}\Big] u_{ B}(p,s), \notag \\
&&\langle B^{\prime}(p',s')|A^{\mu}| B (p,s)\rangle = \bar u_{ B^{\prime}}(p',s') \Big[G_1(q^2)\gamma^{\mu}+G_2(q^2)\frac{p^{\mu}}{m_{ B}}+G_3(q^2)\frac{p'^{\mu}}{m_{B^{\prime}}}\Big]
\gamma_5 u_{ B}(p,s), \notag \\
\end{eqnarray}
where \( F_1(q^2) \), \( F_2(q^2) \), and \( F_3(q^2) \) represent the form factors associated with the vector transition, while \( G_1(q^2) \), \( G_2(q^2) \), and \( G_3(q^2) \) correspond to the form factors of the axial transition. The Dirac spinors for the initial and final baryon states are denoted as \( u_{B}(p, s) \) and \( u_{B^{\prime}}(p', s') \) respectively, and the transferred momentum of the leptons is given by \( q = p - p' \). To compute the form factors, we will employ the three-point correlation function:
\begin{eqnarray}\label{CorFunc}
\Pi_{\mu}(p,p^{\prime},q)&=&i^2\int d^{4}x e^{-ip\cdot x}\int d^{4}y e^{ip'\cdot y}  \langle 0|{\cal T}\{{\cal J}^{B^{\prime}}(y){\cal
J}_{\mu}^{tr,V(A)}(0) \bar {\cal J}^{ B}(x)\}|0\rangle,
\end{eqnarray}
where  we denote the interpolating currents of the initial and final baryons as $ {\cal J}^{ B}(x)$ and ${\cal J}^{B^{\prime}}(y) $, respectively, and the  superscript  $ "tr" $ in transition interpolating current ${\cal
	J}_{\mu}^{tr,V(A)}(0)$ denotes "transition", while \( \cal T \) represents the time-ordering operator. To compute the correlation function of the physical side, we must insert two appropriate complete sets of hadronic states that share the same quantum numbers as the currents $ {\cal J}^{ B}(x)$ and ${\cal J}^{B^{\prime}}(y) $ at the relevant positions. Through a series of algebraic manipulations, we can express the physical side of the correlation function as follows:
\begin{eqnarray} \label{PhysSide}
\Pi_{\mu}^{Phys.}(p,p',q)=\frac{\langle 0 \mid {\cal J}^{B^{\prime}} (0)\mid B^{\prime}(p') \rangle \langle B^{\prime} (p')\mid
{\cal J}_{\mu}^{tr,V(A)}(0)\mid B(p) \rangle \langle  B(p)
\mid \bar {\cal J}^{ B}(0)\mid
0\rangle}{(p'^2-m_{B^{\prime}}^2)(p^2-m_{ B}^2)}+\cdots~,
\end{eqnarray}
where $\cdots$ denotes the continuum and higher states. Additionally, we describe the residues for the initial baryons ($\lambda_{ B}$) and the final baryons ($\lambda_{B^{\prime}}$) as follows:

\begin{eqnarray}\label{MatrixElements}
&&\langle 0|{\cal J}^{B^{\prime}}(0)|B^{\prime}(p')\rangle =
\lambda_{B^{\prime}} u_{B^{\prime}}(p',s'), \notag \\
&&\langle B(p)|\bar {\cal J}^{B}(0)| 0 \rangle =
\lambda^{+}_{ B}\bar u_{ B}(p,s).
\end{eqnarray}
 After performing the summations on Dirac spinors (spin 1/2):
\begin{eqnarray}\label{Spinors}
\sum_{s} u_{ B}(p,s)~\bar{u}_{ B}(p,s)&=&\slashed
p+m_{ B},
\end{eqnarray}
we achieve a definition for the physical side. In the following, we utilize the double Borel transformation \cite{Aliev:2006gk}:

\begin{eqnarray}\label{BorelQCD2}
	\mathbf{\widehat{B}}\frac{1}{(p^{2}-s)^m} \frac{1}{(p'^{2}-s')^n}\longrightarrow (-1)^{m+n}\frac{1}{\Gamma[m]\Gamma[n]} \frac{1}{(M^2)^{m-1}}\frac{1}{(M'^2)^{n-1}}e^{-s/M^2} e^{-s'/M'^2},
\end{eqnarray}
where we define the Borel mass parameters \( M^2 \) and \( M'^{2} \), which will be detailed in the subsequent numerical evaluations. The Borel transformation serves to suppress the influence of excited states and continuum contributions, thereby amplifying the ground state effects in both the initial and final states. After implementing the double Borel transformation, we obtain:
\begin{eqnarray}\label{Physical Side structures}
&&\mathbf{\widehat{B}}~\Pi_{\mu}^{Phys.}(p,p',q)=\lambda_{ B}\lambda_{B^{\prime}}~e^{-\frac{m_{ B}^2}{M^2}}
~e^{-\frac{m_{B^{\prime}}^2}{M'^{2}}}\Bigg[F_{1}\bigg(m_{B} m_{B^{\prime}} \gamma_{\mu}+m_{ B} \slashed{p}' \gamma_{\mu}+m_{B^{\prime}}\gamma_{\mu}\slashed {p}+\slashed {p}'\gamma_\mu\slashed {p}\bigg)+\notag\\
&&F_2\bigg(\frac{m_{B^{\prime}}}{m_{B}}p_\mu\slashed {p}+\frac{1}{m_{ B}}p_{\mu}\slashed {p}' \slashed {p}+m_{B^{\prime}}p_\mu +p_\mu\slashed {p}'\bigg)+ F_3\bigg(\frac{1}{m_{B^{\prime}}} p'_{\mu} \slashed {p}' \slashed{p}+p'_\mu\slashed {p}'+p'_\mu\slashed {p}+m_{ B}p'_\mu\bigg)-\notag\\
&& G_1\bigg(m_{B} m_{B^{\prime}} \gamma_{\mu}\gamma_{5}+m_{ B}\slashed {p}'\gamma_\mu\gamma_5-m_{B^{\prime}}\gamma_\mu\slashed {p}\gamma_5-\slashed {p}'\gamma_\mu\slashed {p}\gamma_5\bigg)- G_2\bigg(p_\mu\slashed {p}'\gamma_5+m_{B^{\prime}}p_\mu\gamma_5-\frac{m_{B^{\prime}}}{m_{ B}}p_\mu\slashed {p}\gamma_5-\frac{1}{m_{ B}} p_{\mu} \slashed {p}' \slashed
{p}\gamma_{5}\bigg)\notag\\
&&-G_3\bigg(\frac{m_{ B}}{m_{B^{\prime}}}p'_\mu\slashed {p}'\gamma_5+m_{ B}p'_\mu\gamma_5-\frac{1}{m_{B^{\prime}}} p'_{\mu}
\slashed {p}'\slashed{p}\gamma_{5}-p'_\mu\slashed {p}\gamma_5\bigg)\Bigg]+\cdots~.
\end{eqnarray}
\\
\subsection{QCD approach}

To develop the QCD or OPE approach, we must evaluate the correlation function presented in Eq. (\ref{CorFunc}) by incorporating the appropriate interpolating currents for the initial and final baryon states. This process involves utilizing the interpolating currents specifically designed for sextet and antitriplet single heavy baryons with spin-parity \( J^P=(\frac{1}{2})^+ \), as outlined in \cite{Aliev:2010yx}:

\begin{eqnarray} \label{Current}
{\cal J}^{(s)}_{B'} &&= - {1\over \sqrt{2}} \epsilon^{abc} \Big\{ \Big( q_1^{aT} 
C Q^b \Big) \gamma_5 q_2^c + \beta \Big( q_1^{aT} C \gamma_5 Q^b \Big) q_2^c -
\Big[\Big( Q^{aT} C q_2^b \Big) \gamma_5 q_1^c + \beta \Big( Q^{aT} C
\gamma_5 q_2^b \Big) q_1^c \Big] \Big\}~, 
\end{eqnarray}
\begin{eqnarray} \label{Current2}
{\cal J}^{(anti-t)}_{B'} &&= {1\over \sqrt{6}} \epsilon^{abc} \Big\{ 2 \Big( q_1^{aT} 
C q_2^b \Big) \gamma_5 Q^c + 2 \beta \Big( q_1^{aT} C \gamma_5 q_2^b \Big) Q^c
+ \Big( q_1^{aT} C Q^b \Big) \gamma_5 q_2^c + \beta \Big(q_1^{aT} C
\gamma_5 Q^b \Big) q_2^c   \notag \\
&&
+ \Big(Q^{aT} C q_2^b \Big) \gamma_5 q_1^c +
\beta \Big(Q^{aT} C \gamma_5 q_2^b \Big) q_1^c \Big\}~,
\end{eqnarray}
and we utilize the interpolating currents for doubly charm (bottom) baryons with spin-parity $J^P=(\frac{1}{2})^+$  as follow  \cite{Aliev:2012ru}:
\begin{eqnarray} \label{Currents}
	{\cal J}^{(s)}_{B}(x)&&= \sqrt{2}~\epsilon_{abc}\Bigg\{ \Big(Q^{aT}(x)Cq^{b}(x)\Big)\gamma_{5}Q^{c}(x) + \beta\Big(Q^{aT}(x)C\gamma_{5}q^{b}(x)\Big)Q^{c}(x) \Bigg\},~
\end{eqnarray}
where the indices \( a, b, \) and \( c \)  represent color values, while \( C \) denotes the charge conjugation operator. The symbols \( Q \) and \( q \) indicate heavy and light quark fields, respectively. We introduce a parameter \( \beta \), which will be determined through numerical analysis; specifically, the value \( \beta = -1 \) aligns with the Ioffe current. Our approach to calculating the QCD side begins with the inclusion of the interpolating currents associated with the initial and final baryons, as well as the transition current, into the correlation function. Utilizing Wick’s theorem, we carry out all possible contractions involving the quark fields, enabling us to reformulate the correlation function in terms of the propagators for both the heavy and light quark components.

 We evaluate this calculation for all decay channels described earlier in this context and present the results for all decay channels, but as an example, we only show the calculations for the semileptonic transition  $ \Omega_{bb}^{-}\rightarrow \Xi^0_{b} {\ell} \bar{\nu_{\ell}}$. The final result, in this step,  for this decay channel is:

\begin{eqnarray} \label{ term}
	&&\Pi^{OPE}_{\mu}=i^2 \int d^4x e^{-ipx}\int d^4y e^{ip'y} \frac{1}{\sqrt{3}} \epsilon_{a'b'c'} \epsilon_{abc}\Bigg\{2 \gamma_5~ S^{c a'}_b(y-x) S'^{a b'}_u(y-x) S^{b i}_s(y) \gamma_\mu (1-\gamma_5) S^{i c'}_b(-x) \gamma_5\notag\\
	&&+2 Tr[S^{a b'}_u(y-x) S'^{i a'}_b(-x) (1-\gamma_5)\gamma_\mu S'^{b i}_s(y)] ~\gamma_5 S^{c c'}_b(y-x)\gamma_5-2
	\beta Tr[S^{b i}_s(y)  \gamma_\mu (1-\gamma_5) S^{i a'}_b(-x)~\gamma_5 S'^{a b'}_u(y-x)] \notag\\
	&&\gamma_5 S^{c c'}_b(y-x)+2 \beta \gamma_5~ S^{c a'}_b(y-x) \gamma_5 S'^{a b'}_u(y-x) S^{b i}_s(y) \gamma_\mu (1-\gamma_5) S^{i c'}_b(-x) - Tr[S'^{b a'}_b(y-x) S^{a b'}_u(y-x)] \gamma_5 S^{c i}_s(y) \notag\\
	&&\gamma_\mu(1-\gamma_5) S^{i c'}_b(-x) ~\gamma_5+\gamma_5~ S^{c i}_s(y) \gamma_\mu (1-\gamma_5) S^{i a'}_b(-x) \gamma_5 S'^{a b'}_u(y-x)  S^{b c'}_b(y-x) -\beta \gamma_5 S^{c i}_s(y) \gamma_\mu (1-\gamma_5) S^{i c'}_b(-x)  \notag\\
	&&Tr[  \gamma_5 S'^{b a'}_b(y-x) S^{a b'}_u(y-x) ]+\beta \gamma_5 S^{c i}_s(y)  \gamma_\mu(1-\gamma_5)   S^{i a'}_b(-x) \gamma_5 S'^{a b'}_u(y-x)~S^{b c'}_b(y-x) + \gamma_5 S^{c b'}_u(y-x) S'^{a a'}_b(y-x)  \notag\\
	&&S^{b i}_s(y) \gamma_\mu (1-\gamma_5) S'^{i c'}_b(-x)  \gamma_5  + \gamma_5 S^{c b'}_u(y-x) S'^{i a'}_b(-x)   (1-\gamma_5)\gamma_\mu S'^{b i}_s(y)  S^{a c'}_b(y-x)  \gamma_5 +\beta \gamma_5 S^{c b'}_u(y-x)  \gamma_5  S'^{a a'}_b(y-x) \notag\\
	&&S^{b i}_s(y)  \gamma_\mu(1-\gamma_5)  ~S^{i c'}_b(-x) +\beta \gamma_5 S^{c b'}_u(y-x)  \gamma_5  S'^{i a'}_b(-x)  (1-\gamma_5) \gamma_\mu S'^{b i}_s(y)  ~S^{a c'}_b(y-x) -2 \beta Tr[S'^{a b'}_u(y-x)  \gamma_5  S^{b i}_s(y)  \gamma_\mu\notag\\
	&& (1-\gamma_5) S^{i a'}_b(-x) ] S^{c c'}_b(y-x)  \gamma_5  + 2 \beta Tr[S'^{b a'}_b(y-x)  \gamma_5  S^{a b'}_u(y-x) ] S^{b i}_s(y)  \gamma_\mu (1-\gamma_5)   ~S^{i c'}_b(-x)  \gamma_5 +2 \beta^2 S^{c a'}_b(y-x)  \gamma_5    \notag\\
	&&S'^{a b'}_u(y-x)  \gamma_5 S^{b i}_s(y)   \gamma_\mu (1-\gamma_5) S^{i c'}_b(-x)  - 2 \beta^2 Tr[S'^{a b'}_u(y-x)  \gamma_5  S'^{i a'}_b(-x)  (1-\gamma_5)  \gamma_\mu   S'^{b i}_s(y)   \gamma_5] S^{c c'}_b(y-x)  \notag\\
	&&- 2 \beta  Tr[S'^{a b'}_u(y-x)  \gamma_5  S^{b a'}_b(y-x)] S^{c i}_s(y) \gamma_\mu   (1-\gamma_5)  S^{i c'}_b(-x)     \gamma_5  -  \beta S^{c i}_s(y)  \gamma_\mu (1-\gamma_5) S^{i a'}_b(-x)   S'^{a b'}_u(y)  \gamma_5   S^{b c'}_b(y-x)    \gamma_5] \notag\\
	&&-  \beta^2  Tr[S^{a b'}_u(y-x)  \gamma_5  S'^{b a'}_b(y-x) \gamma_5  ] S^{c i}_s(y) \gamma_\mu   (1-\gamma_5)  S^{i c'}_b(-x)  + \beta^2 S^{c i}_s(y)  \gamma_\mu (1-\gamma_5) S^{i a'}_b(-x)   \gamma_5 S'^{a b'}_u(y-x)    \gamma_5   \notag\\
	&& S^{b c'}_b(y-x)+  \beta  S^{c b'}_u(y-x)   S'^{a a'}_b(y-x) \gamma_5  S^{b i}_s(y) \gamma_\mu   (1-\gamma_5)  S^{i c'}_b(-x)  \gamma_5  + \beta  S^{c b'}_u(y-x) S'^{i a'}_b(-x)   (1-\gamma_5) \gamma_\mu   S'^{b i}_s(y)    \gamma_5   \notag\\
	&& S^{a c'}_b(y-x)  \gamma_5  +  \beta^2  S^{c b'}_u(y-x)  \gamma_5 S'^{i a'}_b(-x)    (1-\gamma_5)  \gamma_\mu  S'^{b i}_s(y)  \gamma_5   S^{a c'}_b(y-x)  + \beta^2 S^{c b'}_u(y-x) \gamma_5  S'^{a a'}_b(y-x)  \gamma_5 S^{b i}_s(y)   \notag\\
	&& \gamma_\mu (1-\gamma_5)  S^{i c'}_b(-x) \Bigg\},
\end{eqnarray}
where $S'=C S^T C$. We use the expression of the light and heavy quark propagators \cite{Agaev:2020zad}:
\begin{eqnarray}\label{LightProp}
	S_{q}^{ab}(x)&=&i\delta _{ab}\frac{\slashed x}{2\pi ^{2}x^{4}}-\delta _{ab}%
	\frac{m_{q}}{4\pi ^{2}x^{2}}-\delta _{ab}\frac{\langle\overline{q}q\rangle}{12} +i\delta _{ab}\frac{\slashed xm_{q}\langle \overline{q}q\rangle }{48}%
	-\delta _{ab}\frac{x^{2}}{192}\langle \overline{q}g_{}\sigma
	Gq\rangle+
	i\delta _{ab}\frac{x^{2}\slashed xm_{q}}{1152}\langle \overline{q}g_{}\sigma Gq\rangle \notag\\
	&-&i\frac{g_{}G_{ab}^{\alpha \beta }}{32\pi ^{2}x^{2}}\left[ \slashed x{\sigma _{\alpha \beta }+\sigma _{\alpha \beta }}\slashed x\right]-i\delta _{ab}\frac{x^{2}\slashed xg_{}^{2}\langle
		\overline{q}q\rangle ^{2}}{7776} -\delta _{ab}\frac{x^{4}\langle \overline{q}q\rangle \langle
		g_{}^{2}G^{2}\rangle }{27648}+\ldots,
\end{eqnarray}
and
\begin{eqnarray}\label{HeavyProp}
	&&S_{Q}^{ab}(x)=i\int \frac{d^{4}k}{(2\pi )^{4}}e^{-ikx}\Bigg
	\{\frac{\delta_{ab}\left( {\slashed k}+m_{Q}\right) }{k^{2}-m_{Q}^{2}}-\frac{g_{}G_{ab}^{\mu \nu}}{4}\frac{\sigma _{\mu\nu }\left( {%
			\slashed k}+m_{Q}\right) +\left( {\slashed k}+m_{Q}\right) \sigma
		_{\mu\nu}}{(k^{2}-m_{Q}^{2})^{2}} +\frac{g_{}^{2}G^{2}}{12}\delta _{ab}m_{Q}\frac{k^{2}+m_{Q}{\slashed k}}{%
		(k^{2}-m_{Q}^{2})^{4}}\notag\\
	&&+\ldots\Bigg \},
\end{eqnarray}
where \( G_{\mu\nu} \) represents the gluon field strength tensor, with \( G_{\mu\nu}^{ab} = G_A^{\mu\nu} t_A^{ab} \), and \( t_A = \frac{\lambda_A}{2} \) and \( G^2 = G_A^{\mu\nu} G_{A\mu\nu} \). The indices \( A \) range from 1 to 8, corresponding to the Gell-Mann matrices.  In the realm of QCD, we analyze the three-points correlation function within the deep Euclidean region using the OPE. Employing Wilson's OPE framework, we observe that the contributions from light and heavy quark propagators involve operators characterized by distinct mass dimensions. The bare-loop term, which has a mass dimension of \( d = 0 \), results in a perturbative impact. In contrast, nonperturbative corrections emerge from operators with varying dimensions, such as \( d = 3 \) for the quark condensate \( \langle \bar{q}q \rangle \), \( d = 4 \) represented by \( \langle G^2 \rangle \), and \( d = 5 \) exemplified by \( \langle q \sigma G \bar{q} \rangle \). After we substitute the quark propagators into the correlation function, the resulting terms are expressed as follows:
\begin{eqnarray}\label{exampleterm}
	\int d^4k\int d^4k' \int d^4x e^{i(k+k'-p).x}\int d^4y e^{i(-k'+p').y}  \frac{1}{(k^2-m_Q^2)^l(k'^2-m_{Q'}^2)^f[y^2]^n [(y-x)^2]^m}.\notag\\
\end{eqnarray}
Using the identity below,  $(y-x)$ and $y$ appear in the  exponential form \cite{Azizi:2017ubq}:

\begin{eqnarray}\label{intx}
	\frac{1}{[(y)^2]^n}&=&\int\frac{d^Dt}{(2\pi)^D}e^{-it\cdot(y)}~i~(-1)^{n+1}~2^{D-2n}~\pi^{D/2}  \frac{\Gamma(D/2-n)}{\Gamma(n)}\Big(-\frac{1}{t^2}\Big)^{D/2-n},
\end{eqnarray}
and
\begin{eqnarray}\label{int}
	\frac{1}{[(y-x)^2]^m}&=&\int\frac{d^Dt'}{(2\pi)^D}e^{-it'\cdot(y-x)}~i~(-1)^{m+1}~2^{D-2m}~\pi^{D/2}  \frac{\Gamma(D/2-m)}{\Gamma(m)}\Big(-\frac{1}{t'^2}\Big)^{D/2-m}.
\end{eqnarray}
By placement  $x_{\mu}\rightarrow
i\frac{\partial}{\partial p_{\mu}}$ and $y_{\mu}\rightarrow
-i\frac{\partial}{\partial p'_{\mu}}$,  we get for instance:
\begin{align}\label{expyx}
	&\int d^4k\int d^4k' \int d^4x e^{i(k+k'-p+t').x}\int d^4y e^{i(-k'+p'-t-t').y} \int{d^Dt'} \int{d^Dt}  \nonumber\\
	& \frac{1}{(k^2-m_Q^2)^l(k'^2-m_{Q'}^2)^f}\Big(-\frac{1}{t^2}\Big)^{D/2-n}\Big(-\frac{1}{t'^2}\Big)^{D/2-m},
\end{align}
and after performing the Fourier integrals, we have:
\begin{eqnarray}\label{fourer}
	\int d^4x e^{i(k+k'-p+t').x}= (2\pi)^4\delta^4(k+k'-p+t') ,
\end{eqnarray}
and
\begin{eqnarray}\label{forier}
	\int d^4y e^{i(-k'+p'-t-t').y}= (2\pi)^4\delta^4(-k'+p'-t-t').
\end{eqnarray}
The Dirac delta functions simplify the expression by eliminating the integrals over \( k \) and \( k' \). To solve the remaining integrals, we use the Feynman parameterization:

\begin{align}
	\label{eq320 }
	&\frac{1}{A_1^{\alpha_1}A_2^{\alpha_2}A_3^{\alpha_3} A_4^{\alpha_4}}=\frac{\Gamma(\alpha_1+\alpha_2+\alpha_3+\alpha_4)}{\Gamma({\alpha_1})\Gamma({\alpha_2})\Gamma({\alpha_3})\Gamma({\alpha_4})} \int_0^1 \int_0^{1-r}\int_0^{1-r-z}dz \, dr \, db \nonumber\\
	&\frac{r^{\alpha_1-1} \ z^{\alpha_2-1} \, b^{\alpha_3-1}(1-r-z-b)^{\alpha_4-1}}{[rA_1 \ +zA_2 \ +bA_3 \ +(1-r-z-b)A_4]^{\alpha_1+\alpha_2+\alpha_3+\alpha_4}}.
\end{align} 
In the final step, we use the following identity to extract the imaginary parts \cite{Azizi:2017ubq}:
\begin{eqnarray}\label{gamma}
	\Gamma[\frac{D}{2}-n](-\frac{1}{L})^{D/2-n}=\frac{(-1)^{n-1}}{(n-2)!}(-L)^{n-2}ln[-L].
\end{eqnarray}
 By evaluating the correlation function, we take into account perturbative contribution and nonperturbative operators with mass dimensions up to five. For the purposes of this study, we assume the masses of the light quarks are zero. Comparing the results from theoretical predictions and experiments data shows that the effect of this approximation is minimal in many studies. Upon integrating, we express the correlation function in terms of 24 distinct Lorentz structures as presented below:
\begin{eqnarray}\label{Structures}
&&\Pi_{\mu}^{\mathrm{OPE}}(p,p',q)=\Pi^{\mathrm{OPE}}_{\slashed{p}' \gamma_{\mu}\slashed{p}}(p^{2},p'^{2},q^{2})~\slashed{p}' \gamma_{\mu}\slashed{p}+
\Pi^{\mathrm{OPE}}_{p_{\mu} \slashed {p}'\slashed {p}}(p^{2},p'^{2},q^{2})~p_{\mu} \slashed {p}'\slashed {p}+
\Pi^{\mathrm{OPE}}_{p_{\mu}' \slashed {p}'\slashed {p}}(p^{2},p'^{2},q^{2})~p_{\mu}' \slashed {p}'\slashed {p}+\Pi^{\mathrm{OPE}}_{p'_\mu\slashed {p}'\gamma_5}(p^{2},p'^{2},q^{2})\notag\\
&&p'_\mu\slashed {p}'\gamma_5+
\Pi^{\mathrm{OPE}}_{p'_\mu\slashed {p}'\slashed{p}\gamma_5}(p^{2},p'^{2},q^{2})~p'_\mu\slashed {p}'\slashed{p}\gamma_5+
\Pi^{\mathrm{OPE}}_{\slashed {p}'\gamma_\mu\gamma_5}(p^{2},p'^{2},q^{2})~\slashed {p}'\gamma_\mu\gamma_5+
\Pi^{\mathrm{OPE}}_{\slashed {p}'\gamma_\mu\slashed {p}\gamma_5}(p^{2},p'^{2},q^{2})~\slashed {p}'\gamma_\mu\slashed {p}\gamma_5+\Pi^{\mathrm{OPE}}_{p_{\mu} \slashed {p}' \slashed{p}\gamma_{5}}(p^{2},p'^{2},q^{2})\notag\\
&&p_{\mu} \slashed {p}' \slashed{p}\gamma_{5}+
 \Pi^{\mathrm{OPE}}_{\slashed{p}' \gamma_{\mu}}(p^{2},p'^{2},q^{2})~\slashed{p}' \gamma_{\mu}+
\Pi^{\mathrm{OPE}}_{p_\mu\slashed {p}'\gamma_5}(p^{2},p'^{2},q^{2})~p_\mu\slashed {p}'\gamma_5+
\Pi^{\mathrm{OPE}}_{p'_\mu\slashed {p}'}(p^{2},p'^{2},q^{2})~p'_\mu\slashed {p}'+
\Pi^{\mathrm{OPE}}_{p_\mu\slashed {p}'}(p^{2},p'^{2},q^{2})~p_\mu\slashed {p}'+\notag\\
&&\Pi^{\mathrm{OPE}}_{\gamma_\mu\slashed {p}\gamma_5}(p^{2},p'^{2},q^{2})~\gamma_\mu\slashed {p}\gamma_5+
\Pi^{\mathrm{OPE}}_{\gamma_{\mu}}(p^{2},p'^{2},q^{2})~\gamma_{\mu}+
\Pi^{\mathrm{OPE}}_{\gamma_{\mu}\slashed {p}}(p^{2},p'^{2},q^{2})~\gamma_{\mu}\slashed {p}+
\Pi^{\mathrm{OPE}}_{ \gamma_{\mu}\gamma_{5}}(p^{2},p'^{2},q^{2}) ~\gamma_{\mu}\gamma_{5}+\Pi^{\mathrm{OPE}}_{p_\mu\slashed {p}\gamma_5}(p^{2},p'^{2},q^{2})\notag\\
&&p_\mu\slashed {p}\gamma_5+
\Pi^{\mathrm{OPE}}_{p'_\mu\slashed {p}\gamma_5}(p^{2},p'^{2},q^{2})~p'_\mu\slashed {p}\gamma_5+
\Pi^{\mathrm{OPE}}_{p'_\mu\slashed {p}}(p^{2},p'^{2},q^{2})~p'_\mu\slashed {p}+
\Pi^{\mathrm{OPE}}_{p_\mu\slashed {p}}(p^{2},p'^{2},q^{2})~p_\mu\slashed {p}+\Pi^{\mathrm{OPE}}_{p'_\mu}(p^{2},p'^{2},q^{2})~p'_\mu+\notag\\
&&
\Pi^{\mathrm{OPE}}_{p'_\mu\gamma_5}(p^{2},p'^{2},q^{2})~p'_\mu\gamma_5+
\Pi^{\mathrm{OPE}}_{p_\mu}(p^{2},p'^{2},q^{2})~p_\mu+
\Pi^{\mathrm{OPE}}_{p_\mu\gamma_5}(p^{2},p'^{2},q^{2})~p_\mu\gamma_5.
\end{eqnarray}
The invariant functions \(\Pi^{\mathrm{OPE}}_i (p^2, p'^2, q^2) \) (where \( i \) indicates different structures) are characterized using double dispersion integrals:
\begin{eqnarray}\label{PiQCD}
\Pi^{\mathrm{OPE}}_i(p^{2},p'^{2},q^{2})&=&\int_{s_{min}}^{\infty}ds
\int_{s'_{min}}^{\infty}ds'~\frac{\rho
^{\mathrm{OPE}}_i(s,s',q^{2})}{(s-p^{2})(s'-p'^{2})} ,
\end{eqnarray}
where $s_{min}=(m_Q+m_Q)^{2}$, $s'_{min}=m_Q^{2}$ and $\rho_i^{\mathrm{OPE}}(s,s',q^{2})$ represent the spectral densities, given by $\rho_i^{\mathrm{OPE}}(s,s',q^{2})=\frac{1}{\pi}Im\Pi^{\mathrm{OPE}}_i (p^2,p'^2,q^2)$. 
 By employing the quark-hadron duality assumption, the upper limits of the integrals are changed to $s_0$ and $s'_0$, which denote the continuum thresholds for the initial and final baryon states. The spectral densities can be expressed as follows:
\begin{equation} \label{Rhoqcd}
\rho^{\mathrm{OPE}}_i(s,s',q^{2})=\rho_i ^{Pert.}(s,s',q^{2})+\sum_{n=3}^{5}\rho_{i}^{n}(s,s',q^{2}).
\end{equation} 
In this study, we denote the perturbative component of our calculations as \(\rho_i^{Pert.}(s,s',q^{2})\) and  the summation \(\sum_{n=3}^{5}\rho_{i}^{n}(s,s',q^{2})\) that captures all mass dimensions relevant to the nonperturbative aspects of our analysis. This includes contributions from the quark condensate, gluon condensate, and the mixed quark-gluon condensate. Additionally, we implement a double Borel transformation on the QCD side and conduct a continuum subtraction. As a result, we have:
\begin{eqnarray}\label{qcd part2}
&&\Pi^{\mathrm{OPE}}_i (M^2,M'^2,s_0,s'_0,q^2)=\int _{s_{min}}^{s_0} ds\int _{s'_{min}}^{s'_0}ds' e^{-s/M^2} e^{-s'/M'^2}\rho
^{\mathrm{OPE}}_{i}(s,s',q^{2}). \nonumber\\
\end{eqnarray}
We provide a detailed examples  of the elements of $\rho_{i}(s,s^{\prime},q^2)$ for the $\gamma_{\mu} \gamma_5$ structure as a case study in the Appendix.

Subsequently, we evaluate the necessary sum rules to achieve the form factors required for numerical calculations by equating the corresponding coefficients from both the physical and QCD sides for the different Lorentz structures. 
This study outlines the sum rules related to form factors, expressing them through baryon masses and their residues. Additionally, it incorporates QCD parameters, including quark and gluon condensates, as well as the masses of quarks. The analysis also involves auxiliary parameters such as \(M^2\), \(M'^2\), \(s_0\), \(s'_0\), and \(\beta\).

\section{Numerical Analysis }\label{Sec3}

The form factors provide all the essential information needed to specify the decay width of the semileptonic decay. The primary goal in this section is to explore the \( q^2 \) dependence of the form factors across the entire physical range. The masses of the particles involved in our numerical calculations are outlined in Table\ \ref{inputParameter}.
\begin{table}[h!]
\caption{Some input masses used in our numerical calculations.}\label{inputParameter}
\begin{tabular}{|c|c|}
\hline 	\hline 
Parameters                                             &  Values  \\
\hline 
$ m_e $                                                & $ 0.51~\mathrm{MeV}$ \cite{ParticleDataGroup:2020ssz}\\
$ m_\mu $                                              & $105~\mathrm{MeV}$ \cite{ParticleDataGroup:2020ssz}\\
$ m_\tau $                                               & $(1776.09\pm0.007)~\mathrm{MeV}$ \cite{ParticleDataGroup:2020ssz}\\
$ m_c$                                                 & $(1.27\pm0.02)~ \mathrm{GeV}$ \cite{ParticleDataGroup:2020ssz}\\
$ m_b$                                                 & $(4.18\pm0.007)~ \mathrm{GeV}$ \cite{ParticleDataGroup:2020ssz}\\
$ m_{ \Xi_{cc}}$                                       & $ (3.62\pm0.0015)~ \mathrm{GeV}$ \cite{ParticleDataGroup:2020ssz}\\
$ m_{ \Omega_{cc}}$                                       & $ (3.70\pm0.09)~ \mathrm{GeV}$ \cite{ShekariTousi:2024mso}\\
$ m_{ \Xi_{bb}}$                                       & $ (9.97\pm0.19)~ \mathrm{GeV}$ \cite{ShekariTousi:2024mso}\\
$ m_{ \Omega_{bb}}$                                       & $ (9.98\pm0.18)~ \mathrm{GeV}$ \cite{ShekariTousi:2024mso}\\
$ m_{\Xi_{c}} $                                      & $ (2.46\pm0.00023)~\mathrm{GeV}$  \cite{ParticleDataGroup:2020ssz} \\
$ m_{\Xi^{'}_{c}} $                                      & $ (2.46\pm0.00023)~\mathrm{GeV}$  \cite{ParticleDataGroup:2020ssz} \\
$ m_{\Omega_{c}} $                                      & $ (2.69\pm0.0017)~\mathrm{GeV}$  \cite{ParticleDataGroup:2020ssz} \\
$ m_{\Lambda_{c}} $                                      & $ (2.28\pm0.00014)~\mathrm{GeV}$  \cite{ParticleDataGroup:2020ssz} \\
$ m_{\Sigma_{c}} $                                      & $ (2.45\pm0.00014)~\mathrm{GeV}$  \cite{ParticleDataGroup:2020ssz} \\
$ m_{\Xi_{b}} $                                      & $ (5.79\pm0.0006)~\mathrm{GeV}$  \cite{ParticleDataGroup:2020ssz} \\
$ m_{\Xi^{'}_{b}} $                                      & $ (5.93\pm0.0005)~\mathrm{GeV}$  \cite{ParticleDataGroup:2020ssz} \\
$ m_{\Lambda_{b}} $                                      & $ (5.61\pm0.00017)~\mathrm{GeV}$  \cite{ParticleDataGroup:2020ssz} \\
$ m_{\Sigma_{b}} $                                      & $ (5.81\pm0.00025)~\mathrm{GeV}$  \cite{ParticleDataGroup:2020ssz} \\
\hline	\hline 
\end{tabular}
\end{table}
We have show the lifetimes of the doubly heavy baryons  in Table\ \ref{Tab:para_doubly_heavy}.
\begin{table}[!htb]
	\caption{Lifetimes (in units of fs) of doubly
		heavy baryons.}
	\label{Tab:para_doubly_heavy} %
	\begin{tabular}{c|c|c|c|c|c|c}
		\hline \hline
		baryons  & $\Xi_{cc}^{++}$  & $\Xi_{cc}^{+}$  & $\Omega_{cc}^{+}$  &  $\Xi_{bb}^{0}$  & $\Xi_{bb}^{-}$  & $\Omega_{bb}^{-}$ \tabularnewline
		\hline 
		
		lifetimes  & 256~\cite{ParticleDataGroup:2020ssz}  & 33~\cite{ParticleDataGroup:2020ssz}  & $270$ \cite{Kiselev:2001fw}  & $370$ \cite{Karliner:2014gca}  & $370$ \cite{Karliner:2014gca}  & $800$\cite{Kiselev:2001fw}\tabularnewline
		\hline \hline
	\end{tabular}
\end{table}

Also, in Tables \ref{res1},  \ref{res2} and \ref{res3}, the residue values are included for doubly heavy baryons and single charm and bottom baryons, respectively.

\begin{table}[!htb]
	\caption{Residues (in units of $\mathrm{GeV}^3$) of doubly
		heavy baryons.}
	\label{res1} %
	\begin{tabular}{c|c|c|c|c}
		\hline \hline
		Baryons  & $\Xi_{cc}$   & $\Omega_{cc}$  &  $\Xi_{bb}$   & $\Omega_{bb}$ \tabularnewline
		\hline 
		
		Residues  & $0.16\pm0.04$ ~\cite{ShekariTousi:2024mso}  &   $0.17\pm0.04$ \cite{ShekariTousi:2024mso}  & $0.45\pm0.08$ \cite{ShekariTousi:2024mso}  & $0.46\pm0.08$\cite{ShekariTousi:2024mso}\tabularnewline
		\hline \hline
	\end{tabular}
\end{table}

\begin{table}[!htb]
	\caption{Residues (in units of $\mathrm{GeV}^3$) of single
		charm baryons.}
	\label{res2} %
	\begin{tabular}{c|c|c|c|c|c|}
		\hline \hline
		Baryons  & $\Xi_{c}$  & $\Xi_{c}^{'}$  & $\Omega_{c}$  & $\Lambda_{c}$  &$\Sigma_{c}$   \tabularnewline
		\hline 
		
		Residues  & $0.027\pm0.004$~\cite{Wang:2010fq}  & $0.055\pm0.007$~\cite{Wang:2009cr}  &  $0.093\pm0.023$~\cite{Wang:2009cr}   &  $0.022\pm0.008$~\cite{Wang:2010fq}   &  $0.045\pm0.015$~\cite{Wang:2009cr}   \tabularnewline
		\hline \hline
	\end{tabular}
\end{table}

\begin{table}[!htb]
	\caption{Residues (in units of $\mathrm{GeV}^3$) of single
		bottom baryons.}
	\label{res3} %
	\begin{tabular}{c|c|c|c|c|}
		\hline \hline
		Baryons  &  $\Xi_{b}$  & $\Xi_{b}^{'}$  & $\Lambda_{b}$& $\Sigma_{b}$ \tabularnewline
		\hline 
		
		Residues  & $0.032\pm0.009$~\cite{Wang:2010fq}  & $0.079\pm0.020$~\cite{Wang:2009cr}  &  $0.030\pm0.009$~\cite{Wang:2010fq}   &  $0.062\pm0.018$~\cite{Wang:2009cr}     \tabularnewline
	\hline \hline
	\end{tabular}
\end{table}

%%%%%%%%%%%%%%%%%%
For the condensate parameters, we utilize some inputs parameters as~\cite{Belyaev:1982sa,Belyaev:1982cd,Ioffe:2005ym}:
$\langle\bar{q}q\rangle=-(0.24\pm0.01{\rm GeV})^{3}$, $\langle\bar{s}s\rangle=(0.8\pm0.2)\langle\bar{q}q\rangle$,
$\langle\bar{q}g_{s}\sigma Gq\rangle=m_{0}^{2}\langle\bar{q}q\rangle$,
$\langle\bar{s}g_{s}\sigma Gs\rangle=m_{0}^{2}\langle\bar{s}s\rangle$,
$m_{0}^{2}=(0.8\pm0.2)\ {\rm GeV}^{2}$, $\langle\frac{\alpha_{s}GG}{\pi}\rangle=(0.012\pm0.004)\ {\rm GeV}^{4}$ and for   Fermi constant and CKM matrix elements we use~\cite{ParticleDataGroup:2020ssz}:
\begin{align}
	& G_{F}=1.166\times10^{-5}{\rm GeV}^{-2},\nonumber \\
	&|V_{ub}|=0.00357,\nonumber \\
	& |V_{cd}|=0.225,\quad|V_{cs}|=0.974.\label{eq:GFCKM}
\end{align}
The sum rules for the form factors involve five additional auxiliary parameters: the Borel parameters \( M^2 \) and \( M^{\prime 2} \), the continuum thresholds \( s_0 \) and \( s^{\prime}_0 \), and the mixing parameter \( \beta \). Ideally, these physical quantities should show minimal sensitivity to these parameters. For this aim, we identify particular intervals for these parameters to guarantee that the form factors display minimal sensitivity to their variations. These intervals are defined according to fundamental criteria, including a minimal dependence on any auxiliary parameter, pole dominance, and the convergence of the OPE.

The maximum values for the Borel mass parameters \( M^2 \) and \( M^{\prime 2} \) are established to certify that the pole contributions exceed those from higher states and the continuum. Therefore, we require
\begin{equation} \label{PC}
PC=\frac{\Pi^{OPE}(M^2,M'^2,s_0,s'_0)}{\Pi^{OPE}(M^2,M'^2,{\infty},{\infty})}\geq 0.5.
\end{equation}
We determine the minimum values for the Borel mass parameters \( M^2 \) and \( M'^2 \) by analyzing the necessary conditions for the convergence of the OPE. For effective convergence, it is crucial that the contributions from perturbative processes surpass those from nonperturbative effects. Additionally, we must guarantee that the influence of nonperturbative operators decreases with increasing dimensions. To achieve this, we introduce ratio (R) and imploy the following condition:
\begin{equation} \label{PC2}
R(M^2, M'^2)=\frac{\Pi^{{OPE}-dim5}(M^2,M'^2,s_0,s'_0)}{\Pi^{OPE}(M^2,M'^2,s_0,s'_0)}\leq0.05.
\end{equation}

According to these specifications, the effective ranges for the Borel mass parameters are outlined in Table \ref{tab2}. The continuum thresholds \(s_0\) and \(s_0'\) are not chosen arbitrarily and we carefully choose these parameters to eliminate any influence from excited states in our calculations. This involves determining specific thresholds by evaluating the stability of sum rules over particular ranges of the Borel mass parameters \( M^2 \) and \( M^{\prime 2} \). To ensure both stability and physical accuracy, we define \( s_0 \) and \( s'_0 \) to represent the beginning of the continuum in the spectrum, thereby minimizing contributions from higher states and continuum. This selection process leads us to roughly working regions for the continuum thresholds, which are defined in Table\ \ref{tab2} as well. 

\begin{table}[!htbp]
	\addtolength{\tabcolsep}{10pt}
	\begin{center}\begin{tabular}{lcc}\hline \hline 
			Channel &Borel mass (GeV$^2$) & Continuum threshold (GeV$^2$)   \\
			\hline\hline
				$\Lambda_{c}$& $[3-5]$ & $[6.60-7.60]$   \\
			$\Xi_{c}$,$\Sigma_{c}$  & $[3-5]$ & $[7.60-8.70]$   \\
						$\Xi'_{c}$& $[3-5]$ & $[8.20-9.40]$   \\
			$\Omega_{c}$ & $[3-5]$ & $[9.00-10.20]$   \\
				$\Lambda_{b}$ & $[5.5-6.5]$ & $[32.60-36.60]$   \\
					$\Xi_{b}$,$\Sigma_{b}$  & $[5.5-6.5]$ & $[35.00-39.00]$   \\
						$\Xi'_{b}$& $[5.5-6.5]$ & $[36.60-40.60]$   \\
				$\Xi_{cc}$, $\Omega_{cc}$ & $[4-6]$ & $[15.00-18.00]$  \\
			$\Xi_{bb}$, $\Omega_{bb}$ & $[10-15]$ & $[112.00-118.00]$   \\
			\hline\hline
		\end{tabular}
	\end{center}
	\caption{Working regions of the Borel mass parameters, ${M}^2$ or $M'^2$, and the continuum thresholds, $s_0$ or $s'_0$, for various channels.}
	\label{tab2}
\end{table}

\begin{figure}[h!] 
	\includegraphics[totalheight=6cm,width=7cm]{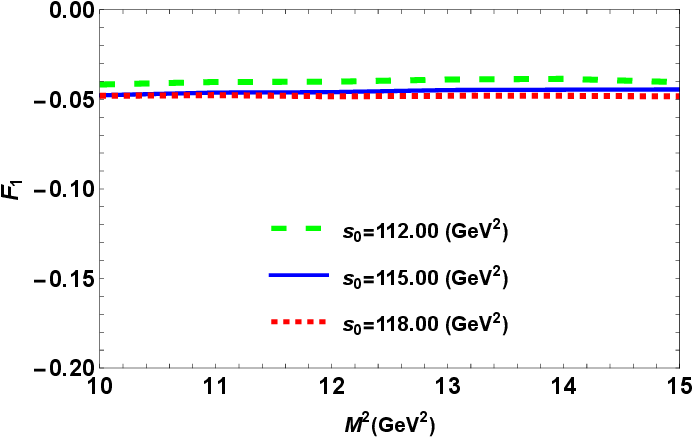}
	\includegraphics[totalheight=6cm,width=7cm]{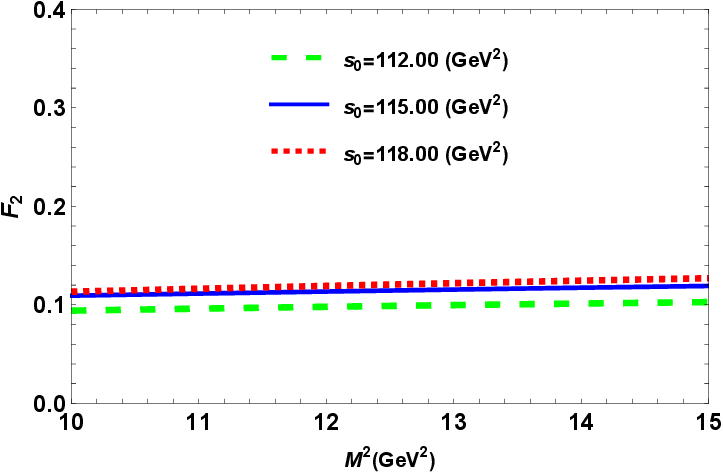}
	\includegraphics[totalheight=6cm,width=7cm]{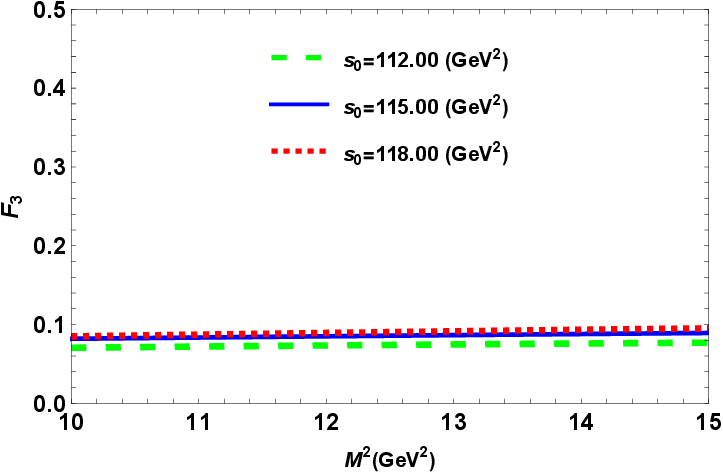}
	\includegraphics[totalheight=6cm,width=7cm]{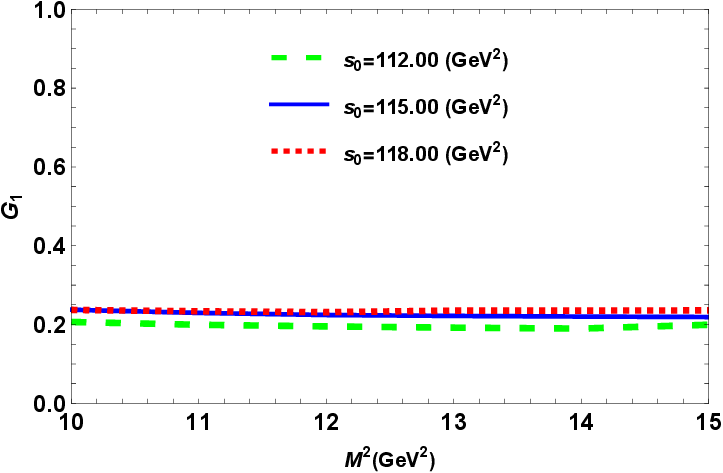}
	\includegraphics[totalheight=6cm,width=7cm]{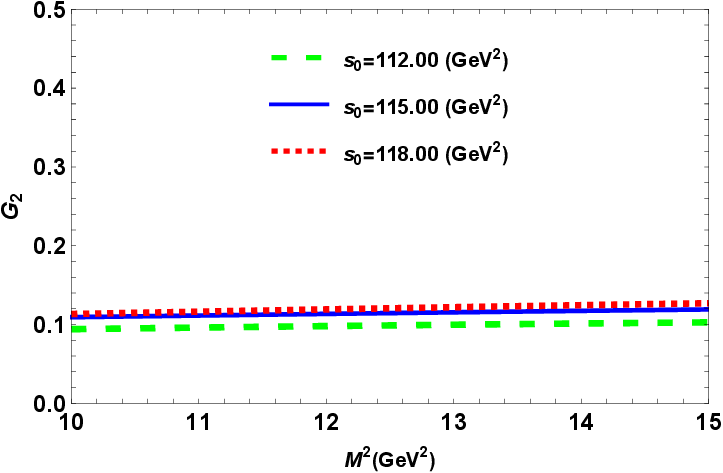}
	\includegraphics[totalheight=6cm,width=7cm]{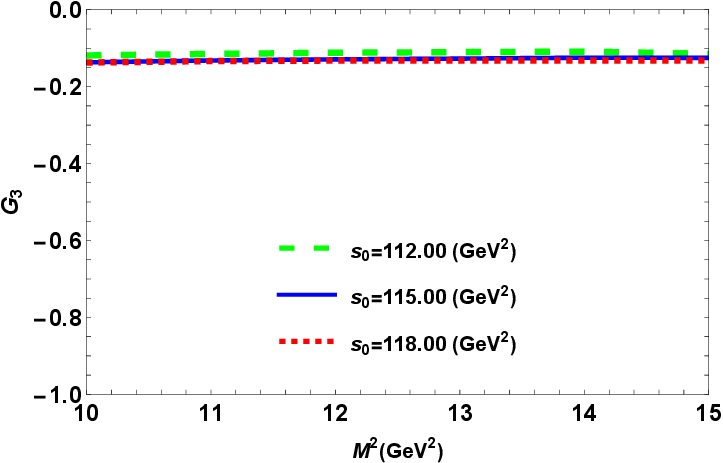}
	\caption{Form factors of $ \Omega_{bb}^{-}\rightarrow \Xi^0_{b} {\ell} \bar{\nu_{\ell}}$,  as an example,  as a function of the Borel parameter \(M^2\) for variuos \(s_0\) amounts, with \(q^2=0\) and other auxiliary parameters set to their central values. The graphs represent the structures \(\slashed{p}'\gamma_{\mu}\), \(p_{\mu}\slashed{p'}\slashed{p}\), \(p'_{\mu}\slashed{p'}\slashed{p}\), \(\gamma_{\mu} \gamma_5\), \(p_{\mu}\slashed{p}'\slashed{p}\gamma_5\), and \(p'_{\mu}\gamma_5\) corresponding to \(F_1\), \(F_2\), \(F_3\), \(G_1\), \(G_2\), and \(G_3\), respectively.}\label{Fig:BorelM}
\end{figure}

\begin{figure}[h!]
	\includegraphics[totalheight=6cm,width=7cm]{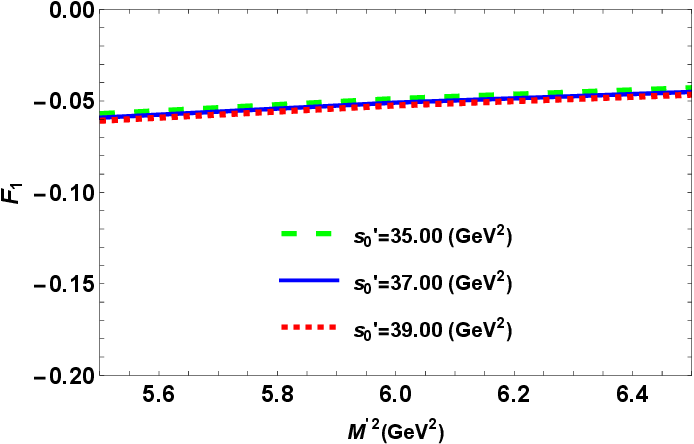}
	\includegraphics[totalheight=6cm,width=7cm]{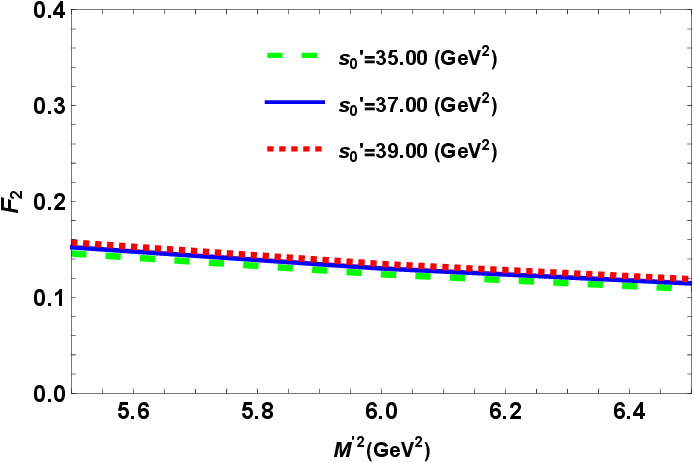}
	\includegraphics[totalheight=6cm,width=7cm]{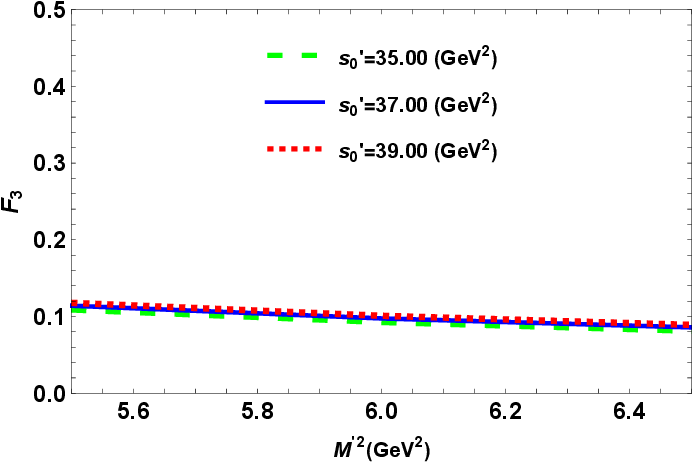}
	\includegraphics[totalheight=6cm,width=7cm]{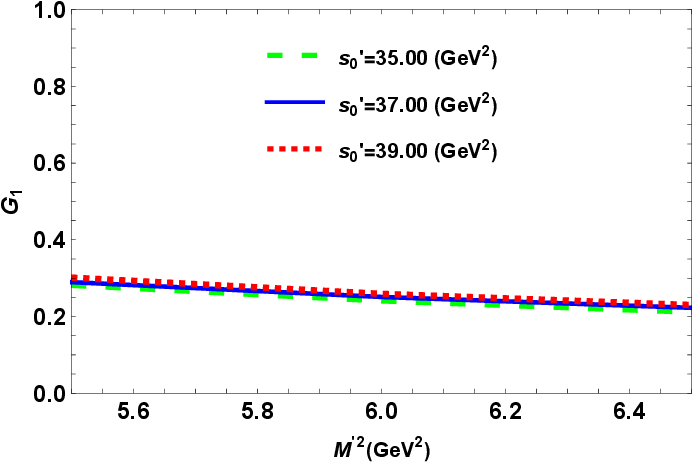}
	\includegraphics[totalheight=6cm,width=7cm]{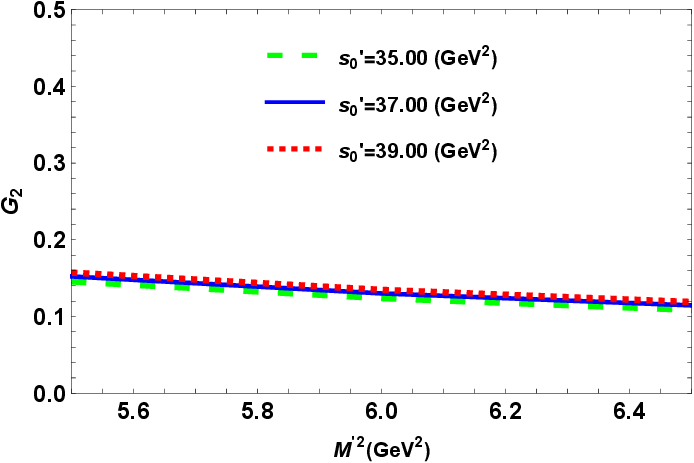}
	\includegraphics[totalheight=6cm,width=7cm]{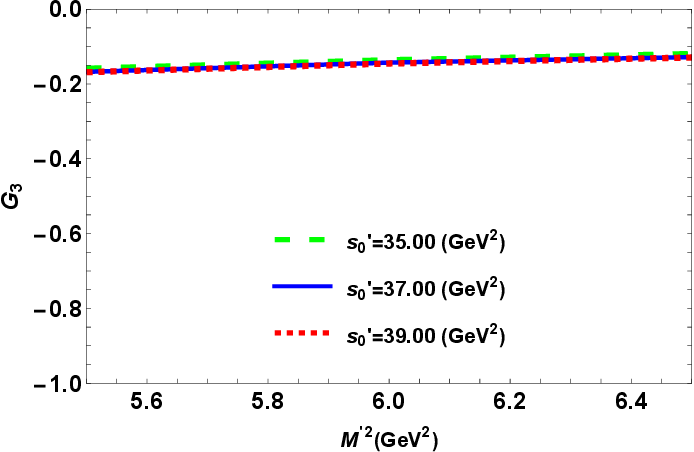}
	\caption{Form factors of $ \Omega_{bb}^{-}\rightarrow \Xi^0_{b} {\ell} \bar{\nu_{\ell}}$,    as an example,  as a function of the Borel parameter \(M'^2\) for variuos \(s'_0\) amounts, with \(q^2=0\) and other auxiliary parameters set to their central values. The graphs represent the structures \(\slashed{p}'\gamma_{\mu}\), \(p_{\mu}\slashed{p'}\slashed{p}\), \(p'_{\mu}\slashed{p'}\slashed{p}\), \(\gamma_{\mu} \gamma_5\), \(p_{\mu}\slashed{p}'\slashed{p}\gamma_5\), and \(p'_{\mu}\gamma_5\) corresponding to \(F_1\), \(F_2\), \(F_3\), \(G_1\), \(G_2\), and \(G_3\), respectively.} \label{Fig:BorelMM}
\end{figure}

As illustrated in Figs. \ref{Fig:BorelM} and \ref{Fig:BorelMM} (note that we present all the figures for only  the transition  $ \Omega_{bb}^{-}\rightarrow \Xi^0_{b} {\ell} \bar{\nu_{\ell}}$,  as an example,  in the present section), the form factors show remarkable stability across variations in the parameters $M^2$, $M'^2$, $s_0$, and $s'_0$ within their individual ranges. This consistency indicates that the parameters have been successfully refined, guaranteeing acceptable outcomes across the specified intervals.

\begin{figure}[h!]
	\includegraphics[totalheight=7cm,width=9cm]{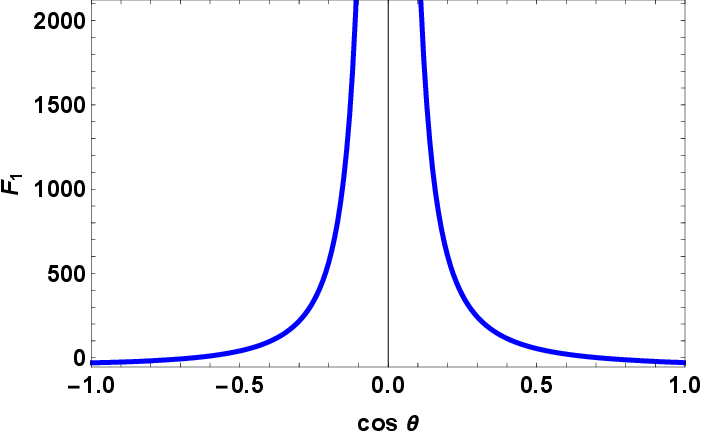}
	\caption{The form factor $F_1$ of $ \Omega_{bb}^{-}\rightarrow \Xi^0_{b} {\ell} \bar{\nu_{\ell}}$ transition,  as an example,   as a function of $\cos \theta$ (or $x$) at $q^2=0$ and average amounts of other auxiliary parameters,  related to the $\slashed{p}'\gamma_{\mu}$ structure. } \label{Fig:f1x}
\end{figure}

The parameters \( M^2 \), \( M'^2 \), \( s_0 \), and \( s'_0 \) hold significance, but the parameter \( \beta \) is also essential. Unlike the others, \( \beta \) has no defined boundaries, allowing it to range from \( -\infty \) to \( \infty \). To determine the most suitable range for \( \beta \), we introduce the transformation \( x = \cos \theta \), where \( \theta = \tan^{-1} \beta \). We choose the range for \( x \) to ensure stability in the form factors, avoiding substantial alterations. For instance, as shown in Fig. \ref{Fig:f1x}, the behavior of the \(F_1\) form factor in relation to \( \cos \theta \) (or \( x \)) is illustrated. Consequently, we limit \( x \) to the regions \( -1 \leq x \leq -0.5 \) and \( 0.5 \leq x \leq 1 \). This restriction applies across all six form factors. Figure \ref{Fig:f1x} highlights that the \(F_1\) form factor remains relatively stable, especially near the Ioffe current at \( x = -0.71 \) within the selected negative interval.

Once we have established the operational ranges for the auxiliary parameters, we move on to explore how the form factors vary with respect to \( q^2 \). By examining the form factors in relation to \( q^2 \), we find that they are well-fit to the subsequent function:
\begin{equation} \label{fitffunction}
{\cal F}(q^2)=\frac{{\cal
F}(0)}{\displaystyle\left(1-a_1\frac{q^2}{m^2_{ B}}+a_2
\frac{q^4}{m_{ B}^4}+a_3\frac{q^6}{m_{B}^6}+a_4\frac{q^8}{m_{ B}^8}\right)}.
\end{equation}

Tables \ref{Tab:parameterfit1} to \ref{Tab:parameterfit14} present the parameters ${\cal F}(0)$, $a_1$, $a_2$, $a_3$, and $a_4$ derived from the central values of the auxiliary parameters for all decay channels. These parameters correspond to six form factors: \(F_1\), \(F_2\), \(F_3\), \(G_1\), \(G_2\), and \(G_3\), which are related  to the structures $\slashed {p}'\gamma_{\mu}  $, $p_{\mu}\slashed {p'}\slashed {p}$, $p'_{\mu}\slashed {p'}\slashed {p}$, $\gamma_{\mu} \gamma_5$, $p_{\mu}\slashed {p}'\slashed {p}\gamma_5$ and $p'_{\mu}\gamma_5$, respectively.

\begin{table}[h!]
\caption{Fit function parameters of different form factors for $ \Xi^{++}_{cc}\rightarrow \Lambda^+_{c} \bar{\ell}\nu_{\ell}$ transition.}\label{Tab:parameterfit1}
\begin{ruledtabular}
\begin{tabular}{|c|c|c|c|c|c|c|}
       Parameter     & $F_1(q^2)$ & $F_2(q^2)$  & $F_3(q^2)$   & $G_1(q^2)$ & $G_2(q^2)$  & $G_3(q^2)$       \\
\hline
${\cal F}(q^2=0)$ & $-0.31\pm0.11$        & $1.24\pm0.40$      & $0.10\pm0.03$     & $0.16\pm0.04$  & $1.23\pm0.33$  & $-1.55\pm0.55$  \\
$a_1$           & $2.63\pm0.81$          & $-0.05\pm0.01$            & $2.99\pm0.98$           & $2.97\pm0.78$           & $2.06\pm0.55$            &$ 2.18\pm0.67$              \\
$a_2$           & $4.63\pm1.44$         & $-111.38\pm33.31$           & $18.17\pm6.07$             & $35.71\pm9.64$          & $-4.36\pm1.01$           & $-22.28\pm7.16$           \\
$a_3$           & $36.98\pm10.91$            & $1329.88\pm380.11$           & $-440.35\pm133.25$          & $-432.60\pm106.32$          & $-448.83\pm112.42$           & $321.21\pm109.23$           \\
$a_4$           & $-658.06\pm221.44$           & $-5206.92\pm1543.99$           & $2300.78\pm777.94$            & $1401.78\pm377.11$         & $2947.41\pm787.14$          & $-1972.39\pm678.46$           \\
\end{tabular}
\end{ruledtabular}
\end{table}
%%%%%%%%%%%%%%
%

\begin{table}[h!]
	\caption{Fit function parameters of different form factors for $ \Xi^{+}_{cc}\rightarrow \Xi^0_{c} \bar{\ell}\nu_{\ell}$ transition.}\label{Tab:parameterfit2}
	\begin{ruledtabular}
		\begin{tabular}{|c|c|c|c|c|c|c|}
		Parameter	& $F_1(q^2)$ & $F_2(q^2)$  & $F_3(q^2)$   & $G_1(q^2)$ & $G_2(q^2)$  & $G_3(q^2)$       \\
			\hline
			${\cal F}(q^2=0)$ & $-0.37\pm0.13$        & $1.35\pm0.43$      & $0.14\pm0.06$     & $0.20\pm0.06$  & $1.34\pm0.43$  & $-1.82\pm0.65$  \\
			$a_1$           & $1.54\pm0.53$          & $-0.05\pm0.01$            & $1.99\pm8.48$           & $2.79\pm0.89$           & $1.07\pm0.26$            &$ 2.15\pm0.72$              \\
			$a_2$           & $-23.84\pm7.45$         & $-111.38\pm32.01$           & $17.17\pm6.63$             & $35.82\pm9.41$          & $-78.83\pm20.83$           & $-27.91\pm8.14$           \\
			$a_3$           & $268.50\pm86.22$            & $1329.88\pm324.11$           & $-442.35\pm160.46$          & $-541.92\pm164.43$          & $935.73\pm239.22$           & $449.20\pm142.20$           \\
			$a_4$           & $-1174.79\pm378.11$           & $-5206.92\pm1544.44$           & $2372.78\pm747.13$            & $2240.25\pm721.09$         & $-3550.64\pm889.32$          & $-2702.99\pm900.13$           \\
		\end{tabular}
	\end{ruledtabular}
\end{table}
%%%%%%%%%%%%%%

\begin{table}[h!]
	\caption{Fit function parameters of different form factors for $ \Omega^{+}_{cc}\rightarrow \Xi^0_{c} \bar{\ell}\nu_{\ell}$ transition.}\label{Tab:parameterfit3}
	\begin{ruledtabular}
		\begin{tabular}{|c|c|c|c|c|c|c|}
			Parameter     & $F_1(q^2)$ & $F_2(q^2)$  & $F_3(q^2)$   & $G_1(q^2)$ & $G_2(q^2)$  & $G_3(q^2)$       \\
			\hline
			${\cal F}(q^2=0)$ & $-0.33\pm0.12$        & $1.41\pm0.50$      & $0.12\pm0.04$     & $0.17\pm0.04$  & $1.41\pm0.33$  & $-1.66\pm0.55$  \\
			$a_1$           & $2.79\pm0.91$          & $0.15\pm0.05$            & $10.14\pm3.50$           & $3.16\pm1.00$           & $-0.91\pm0.22$            &$ 2.31\pm0.82$              \\
			$a_2$           & $5.22\pm1.66$         & $-119.98\pm41.51$           & $71.54\pm23.20$             & $40.25\pm13.22$          & $-84.24\pm21.23$           & $-25.11\pm8.15$           \\
			$a_3$           & $36.98\pm12.00$            & $1329.88\pm421.87$           & $-440.35\pm166.13$          & $-432.60\pm103.98$          & $-448.83\pm122.50$           & $384.40\pm128.90$           \\
			$a_4$           & $-836.33\pm28.77$           & $-5290.92\pm1701.10$           & $1886.22\pm601.54$            & $1781.08\pm477.19$         & $-2879.86\pm726.22$          & $-2506.09\pm825.21$           \\
		\end{tabular}
	\end{ruledtabular}
\end{table}
%%%%%%%%%%%%%%
%
%

\begin{table}[h!]
	\caption{Fit function parameters of different form factors for $ \Xi^{++}_{cc}\rightarrow \Sigma^+_{c} \bar{\ell}\nu_{\ell}$ transition.}\label{Tab:parameterfit4}
	\begin{ruledtabular}
		\begin{tabular}{|c|c|c|c|c|c|c|}
			Parameter     & $F_1(q^2)$ & $F_2(q^2)$  & $F_3(q^2)$   & $G_1(q^2)$ & $G_2(q^2)$  & $G_3(q^2)$       \\
			\hline
			${\cal F}(q^2=0)$ & $-1.55\pm0.50$        & $0.44\pm0.13$      & $0.54\pm0.21$     & $-1.02\pm0.37$  & $0.44\pm0.13$  & $-1.19\pm0.39$  \\
			$a_1$           & $2.34\pm0.80$          & $2.81\pm0.71$            & $4.46\pm14.42$           & $1.91\pm0.63$           & $2.37\pm0.82$            &$ 2.48\pm0.81$              \\
			$a_2$           & $-21.47\pm7.31$         & $-50.88\pm13.55$           & $-36.46\pm12.32$             & $-24.95\pm8.11$          & $-69.49\pm23.12$           & $-35.56\pm12.56$           \\
			$a_3$           & $326.25\pm110.00$            & $756.81\pm160.81$           & $540.92\pm155.34$          & $424.87\pm133.31$          & $972.75\pm315.63$           & $551.08\pm163.87$           \\
			$a_4$           & $-1892.28\pm621.23$           & $-3130.44\pm788.12$           & $-1827.97\pm621.19$            & $-1989.49\pm633.33$         & $-3861.41\pm1398.30$          & $-3598.57\pm1311.21$           \\
		\end{tabular}
	\end{ruledtabular}
\end{table}
%%%%%%%%%%%%%%
%

\begin{table}[h!]
	\caption{Fit function parameters of different form factors for $ \Xi^{++}_{cc}\rightarrow \Xi'^+_{c} \bar{\ell}\nu_{\ell}$ transition.}\label{Tab:parameterfit5}
	\begin{ruledtabular}
		\begin{tabular}{|c|c|c|c|c|c|c|}
			Parameter     & $F_1(q^2)$ & $F_2(q^2)$  & $F_3(q^2)$   & $G_1(q^2)$ & $G_2(q^2)$  & $G_3(q^2)$       \\
			\hline
			${\cal F}(q^2=0)$ & $-1.67\pm0.60$        & $0.47\pm0.17$      & $0.67\pm0.27$     & $-1.11\pm0.39$  & $0.47\pm0.15$  & $-1.28\pm0.44$  \\
			$a_1$           & $2.83\pm0.97$          & $2.81\pm0.90$            & $5.14\pm1.60$           & $2.32\pm0.81$           & $4.69\pm1.45$            &$ 2.98\pm0.96$              \\
			$a_2$           & $3.49\pm1.30$         & $-48.68\pm16.60$           & $1.87\pm0.65$             & $-2.23\pm0.79$          & $35.33\pm12.22$           & $-15.14\pm5.64$           \\
			$a_3$           & $-98.02\pm33.37$            & $940.09\pm280.61$           & $124.19\pm44.70$          & $34.22\pm11.70$          & $-953.32\pm301.11$           & $266.54\pm91.09$           \\
			$a_4$           & $488.55\pm155.13$           & $-5210.79\pm1690.11$           & $-603.14\pm230.65$            & $320.73\pm100.99$         & $7132.38\pm2122.00$          & $-2302.21\pm871.10$           \\
		\end{tabular}
	\end{ruledtabular}
\end{table}
%%%%%%%%%%%%%%
%

\begin{table}[h!]
	\caption{Fit function parameters of different form factors for $ \Xi^{+}_{cc}\rightarrow \Sigma^0_{c} \bar{\ell}\nu_{\ell}$ transition.}\label{Tab:parameterfit6}
	\begin{ruledtabular}
		\begin{tabular}{|c|c|c|c|c|c|c|}
			Parameter     & $F_1(q^2)$ & $F_2(q^2)$  & $F_3(q^2)$   & $G_1(q^2)$ & $G_2(q^2)$  & $G_3(q^2)$       \\
\hline
${\cal F}(q^2=0)$ & $-1.55\pm0.50$        & $0.44\pm0.13$      & $0.54\pm0.21$     & $-1.02\pm0.37$  & $0.44\pm0.13$  & $-1.19\pm0.39$  \\
$a_1$           & $2.34\pm0.80$          & $2.81\pm0.71$            & $4.46\pm14.42$           & $1.91\pm0.63$           & $2.37\pm0.82$            &$ 2.48\pm0.81$              \\
$a_2$           & $-21.47\pm7.31$         & $-50.88\pm13.55$           & $-36.46\pm12.32$             & $-24.95\pm8.11$          & $-69.49\pm23.12$           & $-35.56\pm12.56$           \\
$a_3$           & $326.25\pm110.00$            & $756.81\pm160.81$           & $540.92\pm155.34$          & $424.87\pm133.31$          & $972.75\pm315.63$           & $551.08\pm163.87$           \\
$a_4$           & $-1892.28\pm621.23$           & $-3130.44\pm788.12$           & $-1827.97\pm621.19$            & $-1989.49\pm633.33$         & $-3861.41\pm1398.30$          & $-3598.57\pm1311.21$           \\
		\end{tabular}
	\end{ruledtabular}
\end{table}
%%%%%%%%%%%%%%
%

\begin{table}[h!]
	\caption{Fit function parameters of different form factors for $ \Xi^{+}_{cc}\rightarrow \Xi'^0_{c} \bar{\ell}\nu_{\ell}$ transition.}\label{Tab:parameterfit7}
	\begin{ruledtabular}
		\begin{tabular}{|c|c|c|c|c|c|c|}
			Parameter     & $F_1(q^2)$ & $F_2(q^2)$  & $F_3(q^2)$   & $G_1(q^2)$ & $G_2(q^2)$  & $G_3(q^2)$       \\
		\hline
		${\cal F}(q^2=0)$ & $-1.67\pm0.60$        & $0.47\pm0.17$      & $0.67\pm0.27$     & $-1.11\pm0.39$  & $0.47\pm0.15$  & $-1.28\pm0.44$  \\
		$a_1$           & $2.83\pm0.97$          & $2.81\pm0.90$            & $5.14\pm1.60$           & $2.32\pm0.81$           & $4.69\pm1.45$            &$ 2.98\pm0.96$              \\
		$a_2$           & $3.49\pm1.30$         & $-48.68\pm16.60$           & $1.87\pm0.65$             & $-2.23\pm0.79$          & $35.33\pm12.22$           & $-15.14\pm5.64$           \\
		$a_3$           & $-98.02\pm33.37$            & $940.09\pm280.61$           & $124.19\pm44.70$          & $34.22\pm11.70$          & $-953.32\pm301.11$           & $266.54\pm91.09$           \\
		$a_4$           & $488.55\pm155.13$           & $-5210.79\pm1690.11$           & $-603.14\pm230.65$            & $320.73\pm100.99$         & $7132.38\pm2122.00$          & $-2302.21\pm871.10$           \\
		\end{tabular}
	\end{ruledtabular}
\end{table}
%%%%%%%%%%%%%%
%

\begin{table}[h!]
	\caption{Fit function parameters of different form factors for $ \Omega^{+}_{cc}\rightarrow \Xi'^0_{c} \bar{\ell}\nu_{\ell}$ transition.}\label{Tab:parameterfit8}
	\begin{ruledtabular}
		\begin{tabular}{|c|c|c|c|c|c|c|}
			Parameter     & $F_1(q^2)$ & $F_2(q^2)$  & $F_3(q^2)$   & $G_1(q^2)$ & $G_2(q^2)$  & $G_3(q^2)$       \\
			\hline
			${\cal F}(q^2=0)$ & $-1.41\pm0.49$        & $0.46\pm0.13$      & $0.54\pm0.21$     & $-0.90\pm0.34$  & $0.46\pm0.11$  & $-1.08\pm0.32$  \\
			$a_1$           & $2.94\pm0.92$          & $1.09\pm0.38$            & $4.44\pm2.00$           & $3.45\pm1.12$           & $0.77\pm0.20$            &$ 2.88\pm0.72$              \\
			$a_2$           & $-10.29\pm3.55$         & $-69.88\pm22.32$           & $-11.46\pm4.01$             & $-42.55\pm14.11$          & $-89.86\pm23.35$           & $-28.33\pm9.77$           \\
			$a_3$           & $200.95\pm64.56$            & $748.81\pm212.23$           & $313.38\pm130.39$          & $-858.65\pm272.11$          & $1154.75\pm310.32$           & $478.46\pm155.35$           \\
			$a_4$           & $-1542.68\pm500.09$           & $-2826.30\pm890.03$           & $-740.11\pm280.30$            & $4714.66\pm1550.65$         & $-5448.96\pm1331.82$          & $-3660.69\pm1230.55$           \\
		\end{tabular}
	\end{ruledtabular}
\end{table}
%%%%%%%%%%%%%%
%

\begin{table}[h!]
	\caption{Fit function parameters of different form factors for $ \Omega^{+}_{cc}\rightarrow \Omega^0_{c} \bar{\ell}\nu_{\ell}$ transition.}\label{Tab:parameterfit9}
	\begin{ruledtabular}
		\begin{tabular}{|c|c|c|c|c|c|c|}
			Parameter     & $F_1(q^2)$ & $F_2(q^2)$  & $F_3(q^2)$   & $G_1(q^2)$ & $G_2(q^2)$  & $G_3(q^2)$       \\
			\hline
			${\cal F}(q^2=0)$ & $-1.00\pm0.30$        & $0.33\pm0.12$      & $0.42\pm0.17$     & $-0.64\pm0.27$  & $0.33\pm0.11$  & $-0.76\pm0.34$  \\
			$a_1$           & $2.94\pm0.97$          & $4.52\pm1.51$            & $2.52\pm0.80$           & $3.48\pm1.10$           & $87.48\pm27.45$            &$ 2.88\pm1.30$              \\
			$a_2$           & $-10.29\pm3.77$         & $-9.11\pm3.30$           & $-85.48\pm27.22$             & $36.29\pm13.25$          & $-77.20\pm24.44$           & $-28.33\pm12.39$           \\
			$a_3$           & $200.99\pm71.21$            & $-29.44\pm10.10$           & $134.26\pm45.51$          & $-647.83\pm212.33$          &$ 749.89\pm244.50$           & $478.46\pm212.44$           \\
			$a_4$           & $-1542.55\pm510.19$           & $1408.79\pm444.65$           & $5181.55\pm1590.55$            & $4371.73\pm1490.10$         & $-1849.93\pm610.88$          & $-3660.69\pm1550.87$           \\
		\end{tabular}
	\end{ruledtabular}
\end{table}
%%%%%%%%%%%%%%
%
\begin{table}[h!]
	\caption{Fit function parameters of different form factors for $ \Xi^{-}_{bb}\rightarrow \Lambda^0_{b} {\ell} \bar{\nu}_{\ell}$ transition.}\label{Tab:parameterfit10}
	\begin{ruledtabular}
		\begin{tabular}{|c|c|c|c|c|c|c|}
			Parameter     & $F_1(q^2)$ & $F_2(q^2)$  & $F_3(q^2)$   & $G_1(q^2)$ & $G_2(q^2)$  & $G_3(q^2)$       \\
			\hline
			${\cal F}(q^2=0)$ & $-0.04\pm0.02$        & $0.10\pm0.04$      & $0.07\pm0.03$     & $0.23\pm0.04$  & $0.10\pm0.05$  & $-0.11\pm0.04$  \\
			$a_1$           & $3.32\pm1.30$          & $9.18\pm4.03$            & $8.70\pm3.89$           & $3.31\pm0.66$           & $9.18\pm4.77$            &$ 1.58\pm0.50$              \\
			$a_2$           & $3.68\pm1.45$         & $34.84\pm14.40$           & $20.67\pm8.92$             & $13.82\pm2.30$          & $34.84\pm15.89$           & $-43.91\pm14.56$           \\
			$a_3$           & $-42.06\pm20.00$            & $-35.80\pm15.68$           & $67.08\pm30.30$          & $-172.13\pm35.99$          & $-35.80\pm17.92$           & $374.18\pm134.11$           \\
			$a_4$           & $98.43\pm46.19$           & $-69.53\pm32.33$           & $-292.34\pm140.44$            & $535.16\pm104.10$         & $-69.53\pm34.80$          & $-1103.82\pm379.05$           \\
		\end{tabular}
	\end{ruledtabular}
\end{table}
%%%%%%%%%%%%%%
%
%
\begin{table}[h!]
	\caption{Fit function parameters of different form factors for $ \Omega^{-}_{bb}\rightarrow \Xi^0_{b} {\ell} \bar{\nu}_{\ell}$ transition.}\label{Tab:parameterfit11}
	\begin{ruledtabular}
		\begin{tabular}{|c|c|c|c|c|c|c|}
			Parameter     & $F_1(q^2)$ & $F_2(q^2)$  & $F_3(q^2)$   & $G_1(q^2)$ & $G_2(q^2)$  & $G_3(q^2)$       \\
			\hline
			${\cal F}(q^2=0)$ & $-0.04\pm0.01$        & $0.11\pm0.05$      & $0.08\pm0.02$     & $0.22\pm0.08$  & $0.11\pm0.03$  & $-0.12\pm0.03$  \\
			$a_1$           & $3.60\pm0.90$          & $8.48\pm4.00$            & $8.00\pm2.01$           & $4.18\pm1.56$           & $8.48\pm2.34$            &$ 1.21\pm0.31$              \\
			$a_2$           & $8.94\pm2.47$         & $18.09\pm9.01$           & $4.22\pm1.05$             & $35.05\pm13.01$          & $18.09\pm4.87$           & $-53.85\pm14.00$           \\
			$a_3$           & $-77.06\pm18.21$            & $98.05\pm48.30$           & $197.69\pm44.30$          & $-364.71\pm123.18$          & $98.05\pm23.02$           & $462.03\pm116.90$           \\
			$a_4$           & $175.13\pm44.77$           & $-420.77\pm211.13$           & $-632.86\pm150.99$            & $-420.77\pm160.10$         & $-1353.90\pm324.70$          & $-1173.87\pm301.30$           \\
		\end{tabular}
	\end{ruledtabular}
\end{table}
%%%%%%%%%%%%%%
%

\begin{table}[h!]
	\caption{Fit function parameters of different form factors for $ \Xi^{0}_{bb}\rightarrow \Sigma^+_{b} {\ell} \bar{\nu}_{\ell}$ transition.}\label{Tab:parameterfit12}
	\begin{ruledtabular}
		\begin{tabular}{|c|c|c|c|c|c|c|}
			Parameter     & $F_1(q^2)$ & $F_2(q^2)$  & $F_3(q^2)$   & $G_1(q^2)$ & $G_2(q^2)$  & $G_3(q^2)$       \\
			\hline
			${\cal F}(q^2=0)$ & $-0.04\pm0.01$        & $0.21\pm0.09$      & $0.014\pm0.005$     & $0.018\pm0.008$  & $0.25\pm0.09$  & $-0.37\pm0.13$  \\
			$a_1$           & $5.47\pm1.53$          & $8.55\pm3.03$            & $7.96\pm2.69$           & $3.98\pm1.22$           & $8.55\pm3.00$            &$ 3.84\pm1.11$              \\
			$a_2$           & $-1.14\pm0.34$         & $20.34\pm9.13$           & $33.15\pm11.19$             & $0.03\pm0.01$          & $20.34\pm7.17$           & $-7.95\pm2.73$           \\
			$a_3$           & $82.00\pm22.22$            & $79.30\pm30.01$           & $-95.63\pm30.10$          & $-3.64\pm1.23$          & $79.30\pm26.04$           & $105.66\pm34.60$           \\
			$a_4$           & $-225.92\pm59.09$           & $-372.25\pm149.02$           & $170.86\pm57.78$            & $39.17\pm13.17$         & $-372.25\pm128.99$          & $-294.28\pm92.49$           \\
		\end{tabular}
	\end{ruledtabular}
\end{table}
%%%%%%%%%%%%%%
%

\begin{table}[h!]
	\caption{Fit function parameters of different form factors for $ \Xi^{-}_{bb}\rightarrow \Sigma^0_{b} {\ell} \bar{\nu}_{\ell}$ transition.}\label{Tab:parameterfit13}
	\begin{ruledtabular}
		\begin{tabular}{|c|c|c|c|c|c|c|}
				Parameter     & $F_1(q^2)$ & $F_2(q^2)$  & $F_3(q^2)$   & $G_1(q^2)$ & $G_2(q^2)$  & $G_3(q^2)$       \\
			\hline
			${\cal F}(q^2=0)$ & $-0.04\pm0.01$        & $0.21\pm0.09$      & $0.014\pm0.005$     & $0.018\pm0.008$  & $0.25\pm0.09$  & $-0.37\pm0.13$  \\
			$a_1$           & $5.47\pm1.53$          & $8.55\pm3.03$            & $7.96\pm2.69$           & $3.98\pm1.22$           & $8.55\pm3.00$            &$ 3.84\pm1.11$              \\
			$a_2$           & $-1.14\pm0.34$         & $20.34\pm9.13$           & $33.15\pm11.19$             & $0.03\pm0.01$          & $20.34\pm7.17$           & $-7.95\pm2.73$           \\
			$a_3$           & $82.00\pm22.22$            & $79.30\pm30.01$           & $-95.63\pm30.10$          & $-3.64\pm1.23$          & $79.30\pm26.04$           & $105.66\pm34.60$           \\
			$a_4$           & $-225.92\pm59.09$           & $-372.25\pm149.02$           & $170.86\pm57.78$            & $39.17\pm13.17$         & $-372.25\pm128.99$          & $-294.28\pm92.49$           \\
		\end{tabular}
	\end{ruledtabular}
\end{table}
%%%%%%%%%%%%%%
%

\begin{table}[h!]
	\caption{Fit function parameters of different form factors for $ \Omega^{-}_{bb}\rightarrow \Xi'^0_{b} {\ell} \bar{\nu}_{\ell}$ transition.}\label{Tab:parameterfit14}
	\begin{ruledtabular}
		\begin{tabular}{|c|c|c|c|c|c|c|}
			Parameter     & $F_1(q^2)$ & $F_2(q^2)$  & $F_3(q^2)$   & $G_1(q^2)$ & $G_2(q^2)$  & $G_3(q^2)$       \\
			\hline
			${\cal F}(q^2=0)$ & $-0.06\pm0.02$        & $0.34\pm0.15$      & $0.02\pm0.01$     & $0.017\pm0.006$  & $0.28\pm0.10$  & $-0.12\pm0.03$  \\
			$a_1$           & $5.45\pm1.80$          & $9.98\pm4.89$            & $10.15\pm4.80$           & $3.88\pm1.30$           & $9.98\pm3.30$            &$ 3.96\pm0.96$              \\
			$a_2$           & $-1.98\pm0.67$         & $55.96\pm24.90$           & $55.55\pm25.38$             & $-3.05\pm1.01$          & $55.96\pm17.98$           & $-4.71\pm1.14$           \\
			$a_3$           & $89.84\pm28.30$            & $-222.90\pm95.90$           & $-145.40\pm67.05$          & $32.45\pm10.70$          & $-222.90\pm71.91$           & $75.46\pm18.04$           \\
			$a_4$           & $-248.79\pm80.98$           & $472.05\pm224.75$           & $81.56\pm37.91$            & $-47.00\pm16.10$         & $472.06\pm150.24$          & $-203.39\pm50.93$           \\
		\end{tabular}
	\end{ruledtabular}
\end{table}
%%%%%%%%%%%%%%
%

QCD sum rule approach proposes various structures for the choosing form factors. By examining the Borel and continuum ranges alongside the $x $ parameter, we can make effective selections that minimize uncertainties in the outcomes.
  As mentioned, the six structures chosen— $\slashed {p}'\gamma_{\mu}  $, $p_{\mu}\slashed {p'}\slashed {p}$, $p'_{\mu}\slashed {p'}\slashed {p}$, $\gamma_{\mu} \gamma_5$, $p_{\mu}\slashed {p}'\slashed {p}\gamma_5$ and $p'_{\mu}\gamma_5$ —are aligned with the desired form factors $F_1$, $F_2$, $F_3$, $G_1$, $G_2$ and $G_3$, respectively.

\begin{figure}[h!] 
 	\includegraphics[totalheight=6cm,width=7cm]{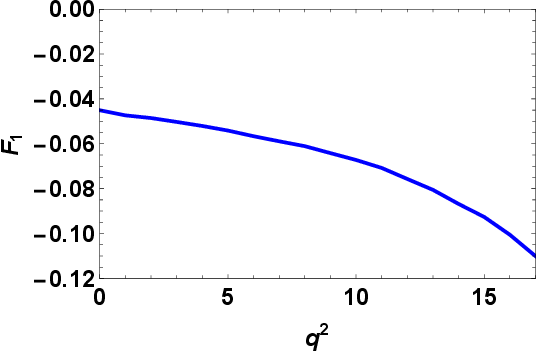}
 	\includegraphics[totalheight=6cm,width=7cm]{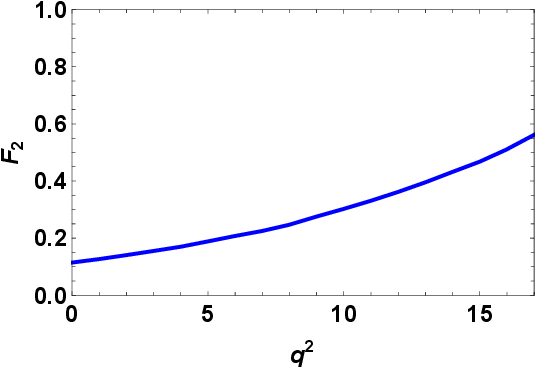}
 	\includegraphics[totalheight=6cm,width=7cm]{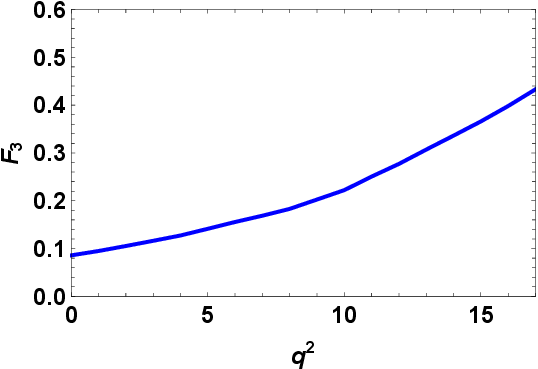}
 	\includegraphics[totalheight=6cm,width=7cm]{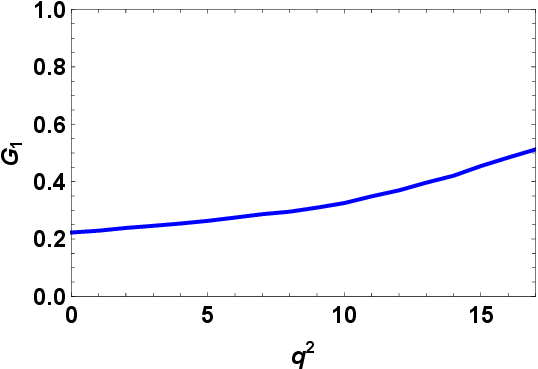}
 	\includegraphics[totalheight=6cm,width=7cm]{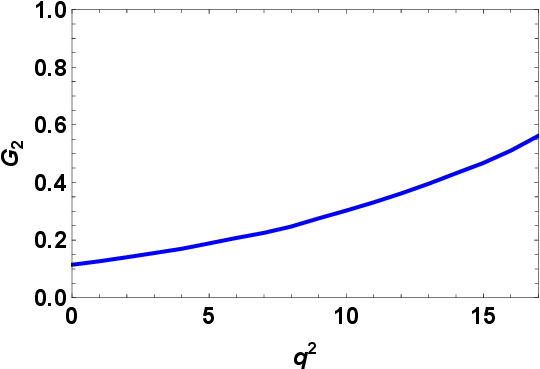}
 	\includegraphics[totalheight=6cm,width=7cm]{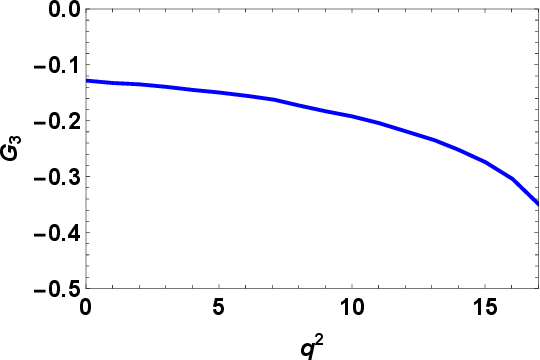}
 	\caption{Form factors of $ \Omega_{bb}^{-}\rightarrow \Xi^0_{b} {\ell} \bar{\nu_{\ell}}$,  as an example,  as  functions of \( q^2 \) at the central values of auxiliary parameters and the Ioffe point  for the chosen structures. }\label{Fig:formfactor1}
 \end{figure}

\begin{figure}[h!] 
 	\includegraphics[totalheight=6cm,width=7cm]{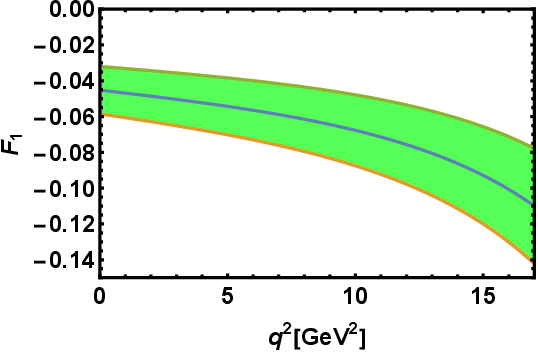}
 	\includegraphics[totalheight=6cm,width=7cm]{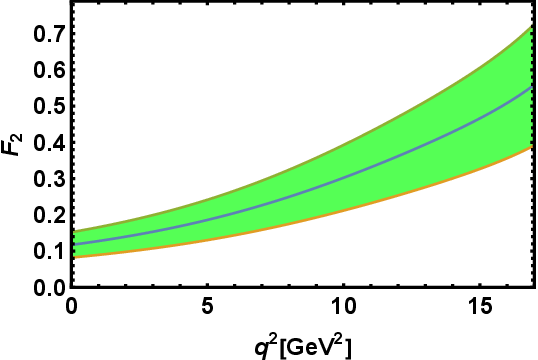}
 	\includegraphics[totalheight=6cm,width=7cm]{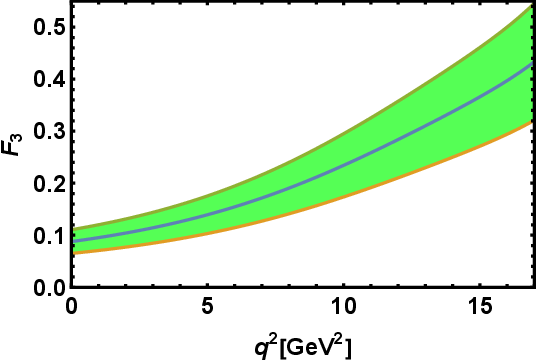}
 	\includegraphics[totalheight=6cm,width=7cm]{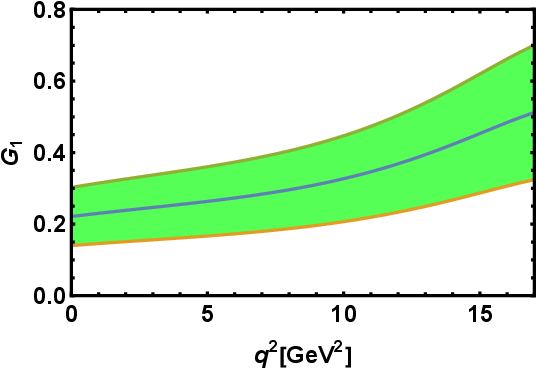}
 	\includegraphics[totalheight=6cm,width=7cm]{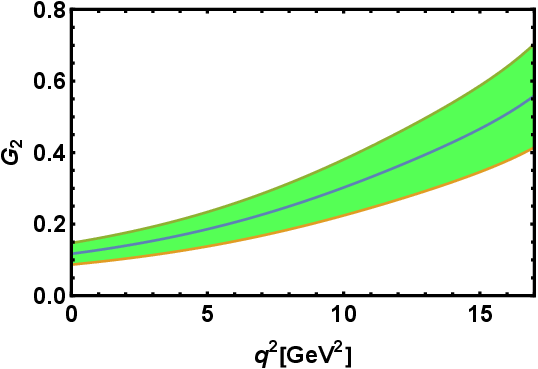}
 	\includegraphics[totalheight=6cm,width=7cm]{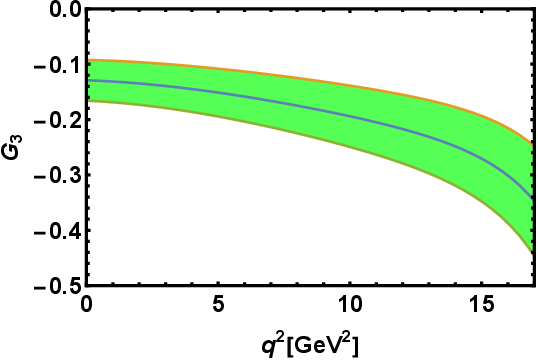}
 	\caption{Form factors and their uncertainties of $ \Omega_{bb}^{-}\rightarrow \Xi^0_{b} {\ell} \bar{\nu_{\ell}}$,  as an example,   as  functions of \( q^2 \) at the central values of auxiliary parameters and the Ioffe point  for the chosen structures.  }\label{Fig:formfactorserror1}
 \end{figure}

 The uncertainties associated with the form factors stem from the variability in auxiliary parameter ranges and uncertainties in other input values. In Figures \ref{Fig:formfactor1} and \ref{Fig:formfactorserror1}, we depict the form factors \( F_1 \), \( F_2 \), \( F_3 \), \( G_1 \), \( G_2 \), and \( G_3 \) as functions of \( q^2 \) at the central values of \( s_0 \), \( s'_0 \), \( M^2 \), \( M'^2 \), and at the Ioffe point \( x = -0.71 \) without and with uncertainties, respectively.

\begin{figure}[h!] 
 	\includegraphics[totalheight=6cm,width=7cm]{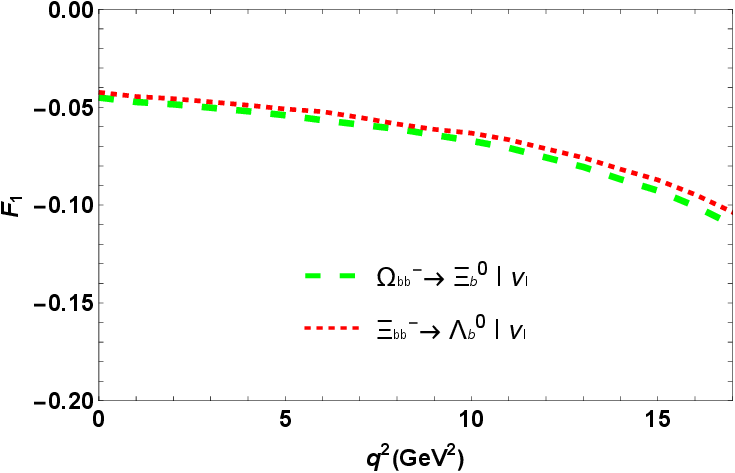}
 	\includegraphics[totalheight=6cm,width=7cm]{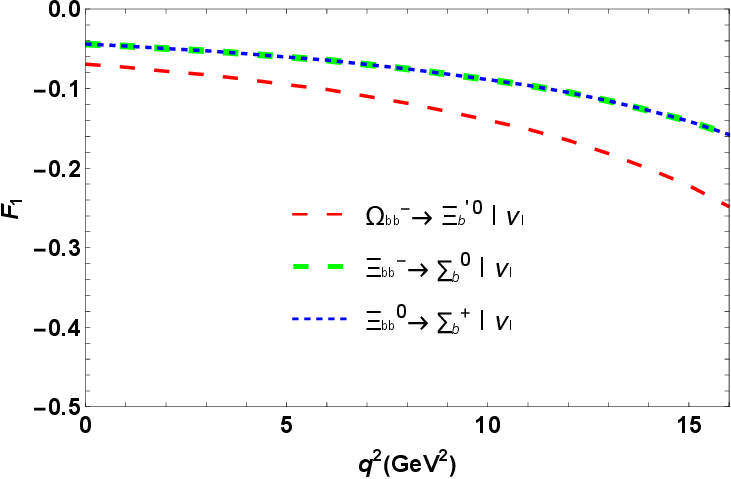}
 	\caption{ Comparision of the \( F_1 \) form factor for the semileptonic decays of triplet to antitriplet (left) and triplet to sextet (right) of the bb sector. }\label{Fig:formfactorf11}
 \end{figure}

  As expected for weak transitions, there is a trend of increasing form factors with higher \( q^2 \). It would also be useful to compare all the form factors of the physical transitions in the same figure for the doubly heavy triplet to single heavy antitriplet/sextet categories in bb sector, as an example.  In Figure \ref{Fig:formfactorf11}, for example, one can see the \( F_1 \) form factor as a function of \( q^2 \) for those transitions.  In the next section, we will apply the fitted expressions for these six form factors within the range \( m_l^2 \leq q^2 \leq (m_{B}-m_{B'})^2 \) to evaluate the decay widths and branching ratios.

%%%%%%%%%%%%%%%%%%%%%%%%%%%%%%%%%%%%%%%%%%%%%%%%%%%%%%%%%%%%%%%%%%%%%%%%%%%%%%%%%%%%%%%%%%%%%%%%%%%%%%%%%%%%
\section{CALCULATION OF Decay Widths and Branching Ratios}~\label{Sec4}
%%%%%%%%%%%%%%%%%%%%%%%%%%%%%%%%%%%%%%%%%%%%%%%%%%%%%%%%%%%%%%%%%

In this section, we define our calculations for key physical observables, including the decay widths and branching ratios related to the semileptonic transitions focusing on all possible lepton channels. 
To do this, we will apply the derived fit functions for the form factors discussed earlier. By using the effective Hamiltonian presented in Eq. (\ref{heff}), we will analyze the decay widths and branching ratios.
 This process incorporates helicity amplitudes for both vector and axial vector currents, which we present below: 

\begin{eqnarray}
	H_{\frac{1}{2},0}^{V} & = & -i\frac{\sqrt{Q_{-}}}{\sqrt{q^{2}}}\left((M_{1}+M_{2})F_{1}-\frac{q^{2}}{M_{1}}F_{2}\right),\;\;\;
	H_{\frac{1}{2},0}^{A} =  -i\frac{\sqrt{Q_{+}}}{\sqrt{q^{2}}}\left((M_{1}-M_{2})G_{1}+\frac{q^{2}}{M_{1}}G_{2}\right),\nonumber \\
	H_{\frac{1}{2},1}^{V} & = & i\sqrt{2Q_{-}}\left(-F_{1}+\frac{M_{1}+M_{2}}{M_{1}}F_{2}\right),\;\;\;
	H_{\frac{1}{2},1}^{A}  =  i\sqrt{2Q_{+}}\left(-G_{1}-\frac{M_{1}-M_{2}}{M_{1}}G_{2}\right),\nonumber \\
	H_{\frac{1}{2},t}^{V} & = & -i\frac{\sqrt{Q_{+}}}{\sqrt{q^{2}}}\left((M_{1}-M_{2})F_{1}+\frac{q^{2}}{M_{1}}F_{3}\right),\;\;\;
	H_{\frac{1}{2},t}^{A} =  -i\frac{\sqrt{Q_{-}}}{\sqrt{q^{2}}}\left((M_{1}+M_{2})G_{1}-\frac{q^{2}}{M_{1}}G_{3}\right),
\end{eqnarray}
where $M_{1}$ and $M_{2}$ represent the masses of the initial and the final baryon, respectively. The variable $Q_{\pm}$ is defined as $Q_{\pm}=(M_1\pm M_2)^{2}-q^{2}$. These helicity amplitudes play a vital role in calculating the decay widths and branching ratios. The amplitudes corresponding to negative helicity are given by:

\begin{equation}
	H_{-\lambda_{2},-\lambda_{W}}^{V}=H_{\lambda_{2},\lambda_{W}}^{V}\quad\text{and}\quad H_{-\lambda_{2},-\lambda_{W}}^{A}=-H_{\lambda_{2},\lambda_{W}}^{A},
\end{equation}
where $\lambda_{2}$ and $\lambda_{W}$ represent the polarizations of the final baryon and the $W$ boson, respectively. Lastly, the helicity amplitudes for the $V-A$ current are detailed as follows:

\begin{equation}
	H_{\lambda_{2},\lambda_{W}}=H_{\lambda_{2},\lambda_{W}}^{V}-H_{\lambda_{2},\lambda_{W}}^{A}.
\end{equation}

The decay width, considering both longitudinally and transversely polarized ${\ell}\nu_{\ell}$ pairs, is calculated as follows:

\begin{align}
	\frac{d\Gamma_{L}}{dq^{2}} & =\frac{G_{F}^{2}|V_{{\rm CKM}}|^{2}q^{2}\ \bold p'\ (1-\hat{m}_{l}^{2})^{2}}{384\pi^{3}M_{1}^{2}}\left((2+\hat{m}_{l}^{2})(|H_{-\frac{1}{2},0}|^{2}+|H_{\frac{1}{2},0}|^{2})+3\hat{m}_{l}^{2}(|H_{-\frac{1}{2},t}|^{2}+|H_{\frac{1}{2},t}|^{2})\right),\label{eq:longi-1}\\
	\frac{d\Gamma_{T}}{dq^{2}} & =\frac{G_{F}^{2}|V_{{\rm CKM}}|^{2}q^{2}\ \bold p'\ (1-\hat{m}_{l}^{2})^{2}(2+\hat{m}_{l}^{2})}{384\pi^{3}M_{1}^{2}}(|H_{\frac{1}{2},1}|^{2}+|H_{-\frac{1}{2},-1}|^{2}),\label{eq:trans-1}
\end{align}

where $\hat{m}_{l}\equiv m_{l}/\sqrt{q^{2}}$, and $\bold p'=\sqrt{Q_{+}Q_{-}}/(2M_{1})$
is the magnitude of three momentum of final baryon in the rest frame of initial baryon. After integrating over the $q^{2}$, we gain the total decay width as follow: 
\begin{equation}
	\Gamma=\int_{m_l^2}^{(M_{1}-M_{2})^{2}}dq^{2}\frac{d\Gamma}{dq^{2}},
\end{equation}
where
\begin{equation}
	\frac{d\Gamma}{dq^{2}}=\frac{d\Gamma_{L}}{dq^{2}}+\frac{d\Gamma_{T}}{dq^{2}}.
\end{equation}

After analyzing the decay width for all decay channels, we present the average values and their uncertainties in Tables \ref{Tab:semi_lep_cc}, \ref{Tab:semi_lepmo_cc}, \ref{Tab:semi_lep_bb}, \ref{Tab:semi_lepmo_bb} and \ref{Tab:semi_lepto_bb}. 

Additionally, we evaluate the branching ratios for all decay channels, and present the obtained results in Tables \ref{Tab:semi_lep_cc}, \ref{Tab:semi_lepmo_cc}, \ref{Tab:semi_lep_bb}, \ref{Tab:semi_lepmo_bb} and \ref{Tab:semi_lepto_bb}.  It’s worth noting that our results exhibit relatively small uncertainties, indicating a successful optimization regarding variations in auxiliary parameters. These findings clear the way for verification through future experimental efforts.

\begin{table}
	\caption{Decay widths and branching ratios results of $cc$ sector  for the $e^+$ channels. It should be noted that the results of the $ \Xi^{++}_{cc}\rightarrow \Xi^+_{c} \bar{\ell}\nu_{\ell}$ decay channel come from our previous work \cite{Tousi:2024usi} for the table consistency.}
	\label{Tab:semi_lep_cc}%
	\begin{tabular}{l|c|c}
		\hline 
		Channel  & $\Gamma/(10^{-14}\ {\rm GeV})$  & ${\cal B}/10^{-3}$   \tabularnewline
		\hline 
		$\Xi_{cc}^{++}\to\Lambda_{c}^{+}e^+ \nu_{e}$  & $1.07^{+0.10}_{-0.12}$  & $4.19\pm0.82$  \tabularnewline
		$\Xi_{cc}^{++}\to\Xi_{c}^{+}e^+ \nu_{e}$  &$7.20^{+0.77}_{-0.59}$&$2.80^{+0.49}_{-0.13}$  \tabularnewline
		$\Xi_{cc}^{+}\to\Xi_{c}^{0}e^+ \nu_{e}$  & $7.20^{+0.77}_{-0.59}$&$4.81^{+0.49}_{-0.43}$  \tabularnewline
		$\Omega_{cc}^{+}\to\Xi_{c}^{0}e^+ \nu_{e}$  & $0.72^{+0.11}_{-0.13}$ & $2.28\pm0.48$  \tabularnewline
		\hline
		$\Xi_{cc}^{++}\to\Sigma_{c}^{+}e^+ \nu_{e}$  & $5.05\pm0.77$  & $19.65\pm1.39$  \tabularnewline
		$\Xi_{cc}^{++}\to\Xi_{c}^{\prime+}e^+ \nu_{e}$  & $63.77^{+4.20}_{-4.22}$  & $248.10^{+11.30}_{-10.23}$  \tabularnewline
		$\Xi_{cc}^{+}\to\Sigma_{c}^{0}e^+ \nu_{e}$  & $4.19\pm0.51$  & $3.37\pm 0.31$ \tabularnewline
		$\Xi_{cc}^{+}\to\Xi_{c}^{\prime0}e^+ \nu_{e}$  & $63.77^{+4.20}_{-4.22}$  & $42.64^{+1.30}_{-1.33}$  \tabularnewline
		$\Omega_{cc}^{+}\to\Xi_{c}^{\prime0}e^+ \nu_{e}$  & $4.19\pm0.42$  & $13.13\pm0.86$  \tabularnewline
		$\Omega_{cc}^{+}\to\Omega_{c}^{0}e^+ \nu_{e}$  & $21.86^{+1.10}_{-1.09}$  & $68.46^{+4.30}_{-4.23}$   \tabularnewline
		\hline 
	\end{tabular}
\end{table}

\begin{table}
	\caption{Decay widths and branching ratios results of $cc$ sector  for the $\mu ^+$ channels.  It should be noted that the results of the $ \Xi^{++}_{cc}\rightarrow \Xi^+_{c} \bar{\ell}\nu_{\ell}$ decay channel come from our previous work \cite{Tousi:2024usi} for the table consistency.}
	\label{Tab:semi_lepmo_cc}%
	\begin{tabular}{l|c|c}
		\hline 
		Channel  & $\Gamma/(10^{-14}\ {\rm GeV})$  & ${\cal B}/10^{-3}$   \tabularnewline
		\hline 
		$\Xi_{cc}^{++}\to\Lambda_{c}^{+}\mu ^+  {\nu}_{\mu }$  & $1.06\pm0.11$  & $4.13\pm0.83$  \tabularnewline
		$\Xi_{cc}^{++}\to\Xi_{c}^{+}\mu ^+  {\nu}_{\mu }$  & $7.01^{+0.66}_{-0.35}$    & $2.72^{+0.47}_{-0.13}$  \tabularnewline
		$\Xi_{cc}^{+}\to\Xi_{c}^{0}\mu ^+  {\nu}_{\mu }$  & $7.01^{+0.66}_{-0.35}$    & $4.68^{+0.19}_{-0.13}$   \tabularnewline
		$\Omega_{cc}^{+}\to\Xi_{c}^{0}\mu ^+  {\nu}_{\mu }$  & $0.71^{+0.11}_{-0.12}$  & $2.23\pm0.44$\tabularnewline
		\hline
		$\Xi_{cc}^{++}\to\Sigma_{c}^{+}\mu ^+  {\nu}_{\mu }$  & $4.91\pm0.72$  & $19.10\pm1.13$  \tabularnewline
		$\Xi_{cc}^{++}\to\Xi_{c}^{\prime+}\mu ^+  {\nu}_{\mu }$  & $61.48^{+4.21}_{-4.24}$  & $239.20^{+10.30}_{-11.23}$ \tabularnewline
		$\Xi_{cc}^{+}\to\Sigma_{c}^{0}\mu ^+  {\nu}_{\mu }$  &  $4.07\pm0.50$  & $3.28\pm 0.23$ \tabularnewline
		$\Xi_{cc}^{+}\to\Xi_{c}^{\prime0}\mu ^+  {\nu}_{\mu }$  & $61.48^{+4.21}_{-4.20}$  & $41.11^{+1.30}_{-1.23}$  \tabularnewline
		$\Omega_{cc}^{+}\to\Xi_{c}^{\prime0}\mu ^+  {\nu}_{\mu }$  & $4.07\pm0.40$  & $12.79\pm 0.80$   \tabularnewline
		$\Omega_{cc}^{+}\to\Omega_{c}^{0}\mu ^+  {\nu}_{\mu }$  &  $21.07^{+1.10}_{-1.08}$  & $65.97^{+4.28}_{-4.22}$    \tabularnewline
		\hline 
	\end{tabular}
\end{table}

\begin{table}
	\caption{Decay widths and branching ratios results of $bb$ sector  for the $e$ channels.}
	\label{Tab:semi_lep_bb}%
	\begin{tabular}{l|c|c}
		\hline 
		Channel  & $\Gamma/(10^{-16}\ {\rm GeV})$  & ${\cal B}/10^{-4}$   \tabularnewline
		\hline 
	$\Xi_{bb}^{-}\to\Lambda_{b}^{0}e \bar{\nu}_{e}$  & $2.01\pm0.23$  & $1.13^{+0.16}_{-0.18}$  \tabularnewline
		$\Omega_{bb}^{-}\to\Xi_{b}^{0}e \bar{\nu}_{e}$  & $1.41\pm0.13$  & $1.71^{+0.20}_{-0.21}$    \tabularnewline
	$\Xi_{bb}^{0}\to\Sigma_{b}^{+}e \bar{\nu}_{e}$  & $1.61^{+0.16}_{-0.17}$  & $0.90\pm0.08$  \tabularnewline
		$\Xi_{bb}^{-}\to\Sigma_{b}^{0}e \bar{\nu}_{e}$  & $1.61^{+0.16}_{-0.17}$  & $0.90\pm0.08$\tabularnewline
	$\Omega_{bb}^{-}\to\Xi_{b}^{\prime0}e \bar{\nu}_{e}$  &  $1.83^{+0.16}_{-0.18}$  & $2.22\pm0.22$  \tabularnewline
		\hline 
	\end{tabular}
\end{table}

\begin{table}
	\caption{Decay widths and branching ratios results of $bb$ sector  for the $\mu $ channels.}
	\label{Tab:semi_lepmo_bb}%
	\begin{tabular}{l|c|c}
	\hline 
Channel  & $\Gamma/(10^{-16}\ {\rm GeV})$  & ${\cal B}/10^{-4}$   \tabularnewline
\hline 
$\Xi_{bb}^{-}\to\Lambda_{b}^{0} \mu \bar{\nu}_{\mu}$  & $2.00^{+0.23}_{-0.21}$  & $1.12\pm0.15$  \tabularnewline
$\Omega_{bb}^{-}\to\Xi_{b}^{0}  \mu \bar{\nu}_{\mu}$  &$1.40^{+0.12}_{-0.13}$  & $1.71^{+0.20}_{-0.21}$ \tabularnewline
$\Xi_{bb}^{0}\to\Sigma_{b}^{+}  \mu \bar{\nu}_{\mu}$  & $1.60^{+0.16}_{-0.15}$  & $0.90\pm0.08$   \tabularnewline
$\Xi_{bb}^{-}\to\Sigma_{b}^{0}  \mu \bar{\nu}_{\mu}$  &  $1.60^{+0.16}_{-0.15}$  & $0.90\pm0.08$\tabularnewline
$\Omega_{bb}^{-}\to\Xi_{b}^{\prime0}  \mu \bar{\nu}_{\mu}$  &  $1.82\pm0.17$  & $2.22^{+0.20}_{-0.22}$  \tabularnewline
\hline 
	\end{tabular}
\end{table}

\begin{table}
	\caption{Decay widths and branching ratios results of $bb$ sector  for the $\tau $ channels.}
	\label{Tab:semi_lepto_bb}%
	\begin{tabular}{l|c|c}
		\hline 
		Channel  & $\Gamma/(10^{-16}\ {\rm GeV})$  & ${\cal B}/10^{-4}$   \tabularnewline
		\hline 
		$\Xi_{bb}^{-}\to\Lambda_{b}^{0} \tau \bar{\nu}_{\tau}$  & $1.14\pm0.17$  & $0.64\pm0.10$  \tabularnewline
		$\Omega_{bb}^{-}\to\Xi_{b}^{0}  \tau \bar{\nu}_{\tau}$  &  $0.74\pm0.08$  & $0.90\pm0.11$  \tabularnewline
		$\Xi_{bb}^{0}\to\Sigma_{b}^{+}  \tau \bar{\nu}_{\tau}$  & $1.03\pm0.10$  & $0.58\pm0.04$  \tabularnewline
		$\Xi_{bb}^{-}\to\Sigma_{b}^{0}  \tau \bar{\nu}_{\tau}$  &  $1.03\pm0.10$  & $0.58\pm0.04$\tabularnewline
		$\Omega_{bb}^{-}\to\Xi_{b}^{\prime0}  \tau \bar{\nu}_{\tau}$  &  $1.11^{+0.10}_{-0.13}$  & $1.35^{+0.15}_{-0.17}$  \tabularnewline
		\hline 
	\end{tabular}
\end{table}

Semileptonic decays play an important role in exploring new physics beyond the standard model and it would also be
interesting to predict the ratios for all decay channels of  $bb $ sector,
\begin{equation}\label{Ratio}
	R=\frac{Br[B\rightarrow B'\tau\bar\nu_\tau]}{Br[B\rightarrow B' e/\mu\bar\nu_{e/\mu}]},
\end{equation}
 which can be used to test the lepton universality. In  the standard model,  all three lepton generations couple identically to the W and Z gauge bosons. By estimating the ratio of branching fractions, and comparing these predictions with upcoming experimental results, one can probe lepton flavor universality  and potentially uncover new physics beyond the SM. In Table \ref{tabrr}, one can see the values of the ratios of branching fractions for all decay channels of  $bb $ sector. Note that, in cc sector the transitions in $ \tau $ channel are not allowed because of kinematical considerations.

  \begin{table}
 	\caption{Ratio of branching fractions  for all decay channels of  $bb $ sector.}
 	\label{tabrr}%
 	\begin{tabular}{c|c}
 		\hline 
 	Channel  & Br[B $\rightarrow$ B$'$ $\tau\bar\nu_\tau$]/Br[B $\rightarrow$ B$'$ $ e/\mu $  $\bar\nu_{e/\mu}$]  \tabularnewline
 	\hline 
 	$\Xi_{bb}^{-}\to\Lambda_{b}^{0} l \bar{\nu}_{l}$  & $0.56\pm0.08$   \tabularnewline
 	$\Omega_{bb}^{-}\to\Xi_{b}^{0}  l \bar{\nu}_{l}$  &  $0.52\pm0.07$    \tabularnewline
 	$\Xi_{bb}^{0}\to\Sigma_{b}^{+}  l \bar{\nu}_{l}$  & $0.64\pm0.10$    \tabularnewline
 	$\Xi_{bb}^{-}\to\Sigma_{b}^{0}  l \bar{\nu}_{l}$  & $0.64\pm0.10$  \tabularnewline
 	$\Omega_{bb}^{-}\to\Xi_{b}^{\prime0}  l \bar{\nu}_{l}$  &   $0.60\pm0.09$  \tabularnewline
 		\hline 
 	\end{tabular}
 \end{table}
 
\begin{figure}[h!]
  	\includegraphics[totalheight=4.5cm,width=7cm]{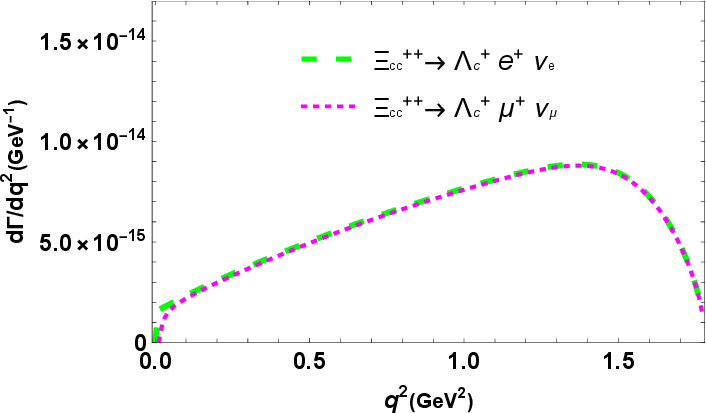}
  	\includegraphics[totalheight=4.5cm,width=7cm]{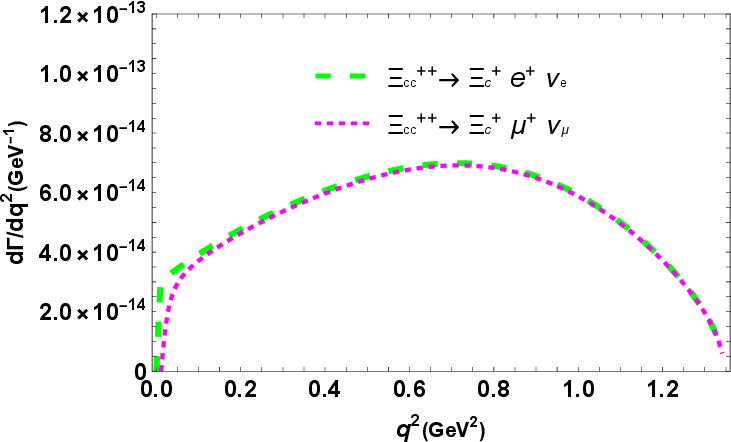}
  	\includegraphics[totalheight=4.5cm,width=7cm]{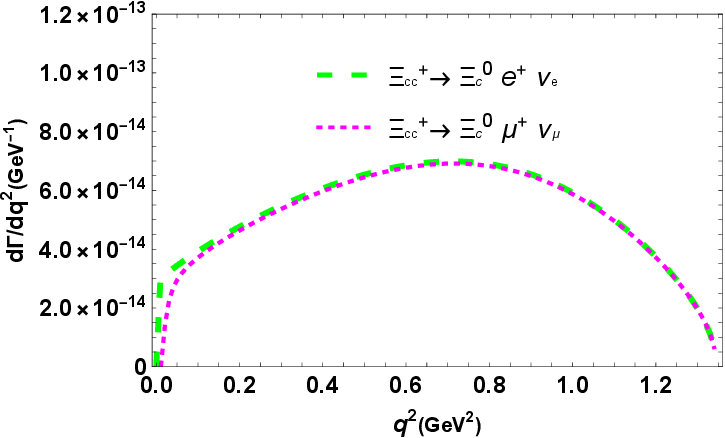}
  	\includegraphics[totalheight=4.5cm,width=7cm]{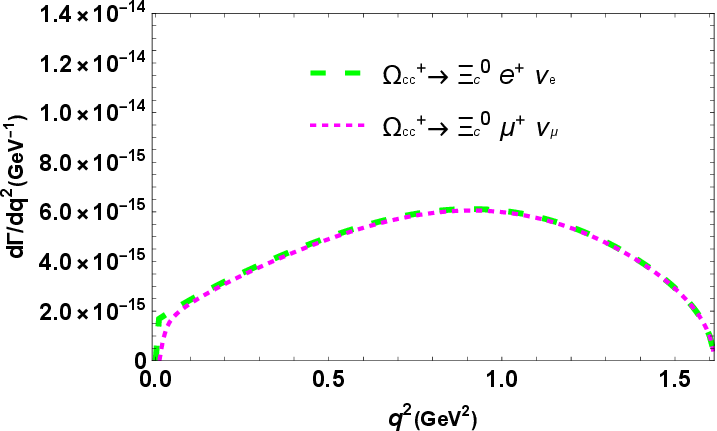}	
  	\includegraphics[totalheight=4.5cm,width=7cm]{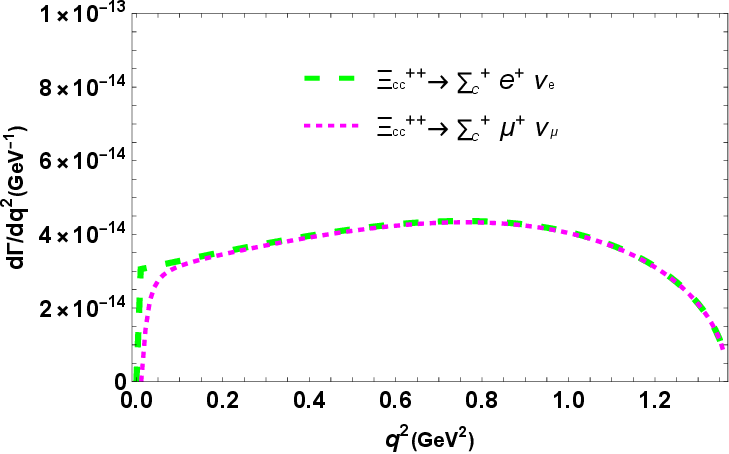}
  	\includegraphics[totalheight=4.5cm,width=7cm]{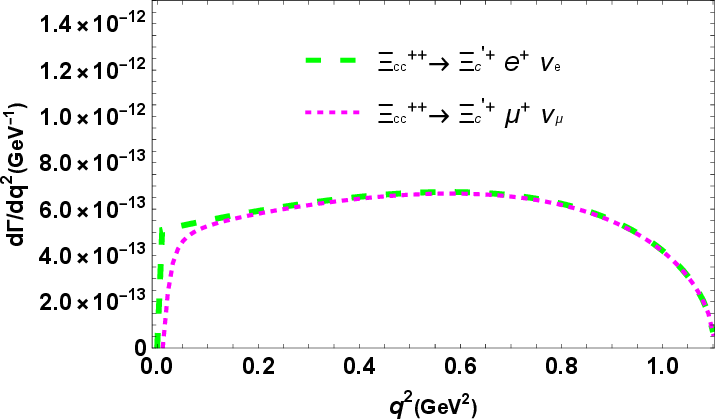}
  	\includegraphics[totalheight=4.5cm,width=7cm]{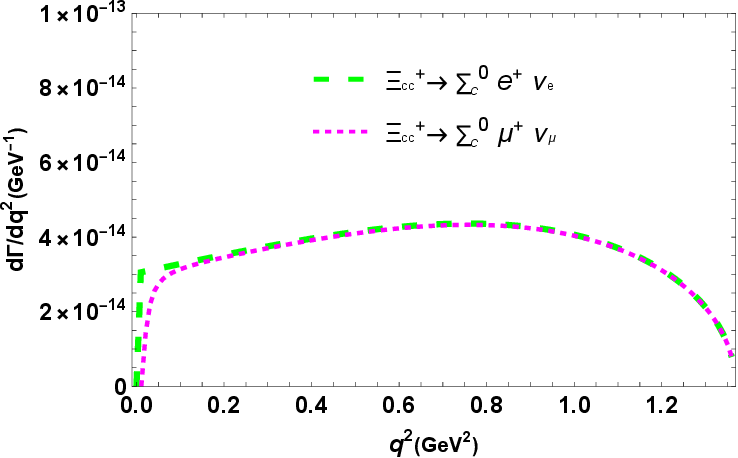}
  	\includegraphics[totalheight=4.5cm,width=7cm]{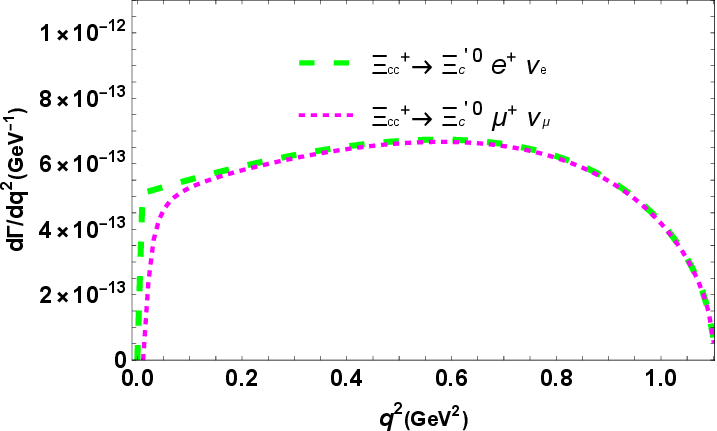}
  	\includegraphics[totalheight=4.5cm,width=7cm]{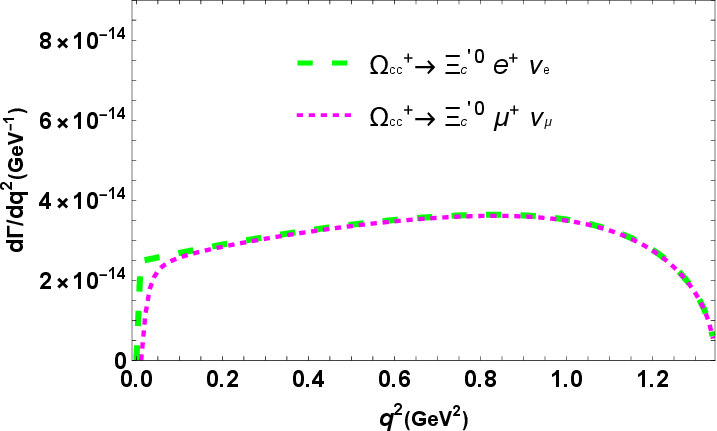}
  	\includegraphics[totalheight=4.5cm,width=7cm]{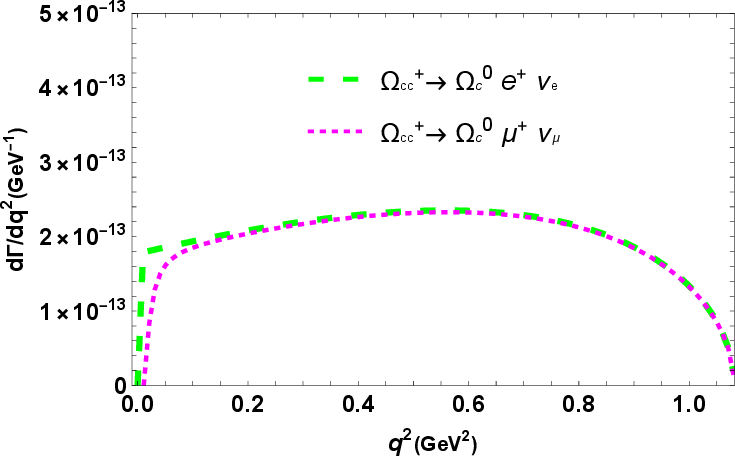}
  	\caption{Diﬀerential decay widths of  $cc$ sector  as  functions of \( q^2 \) for $e$ and $\mu$  channels.} \label{Fig:diff}
  \end{figure}

\begin{figure}[h!]
 	\includegraphics[totalheight=5cm,width=8cm]{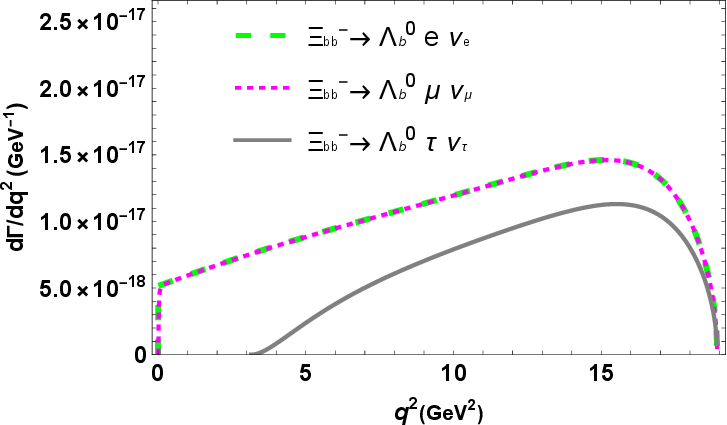}
 	\includegraphics[totalheight=5cm,width=8cm]{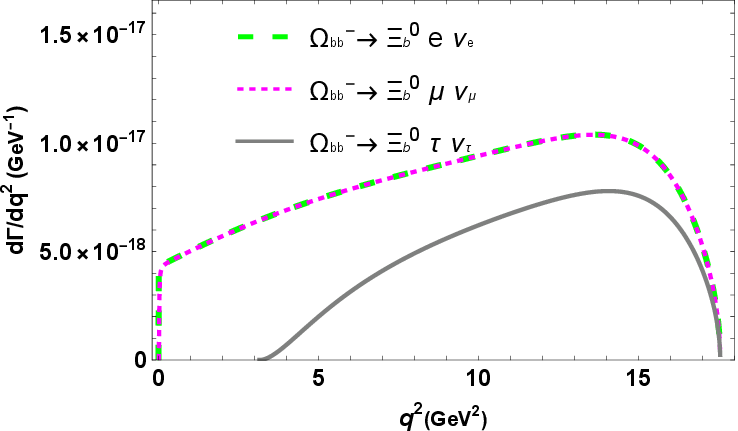}
 	\includegraphics[totalheight=5cm,width=8cm]{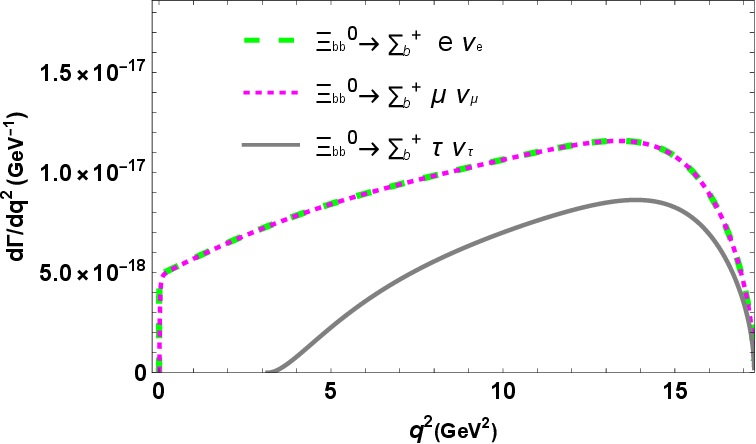}
 	\includegraphics[totalheight=5cm,width=8cm]{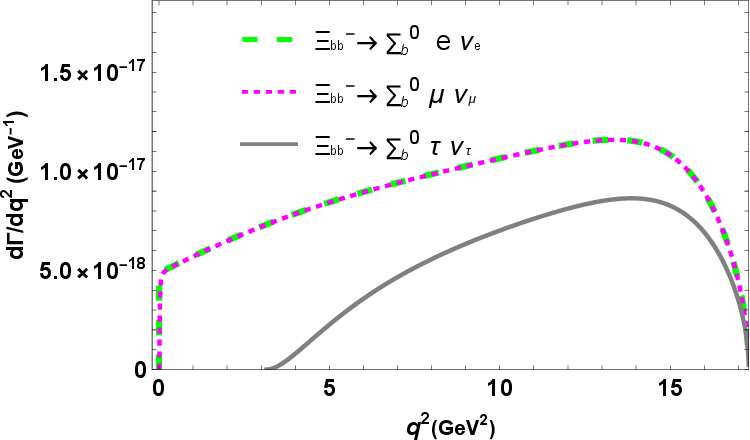}	
 	\includegraphics[totalheight=5cm,width=8cm]{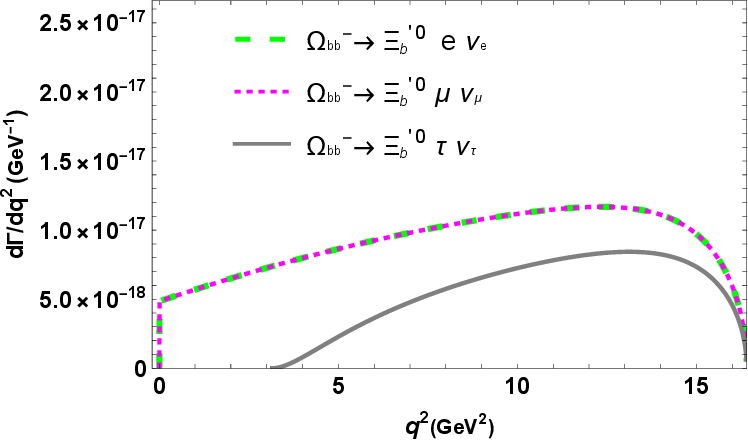}
 	\caption{Diﬀerential decay widths of  $bb$ sector  as  functions of \( q^2 \) for $e$, $\mu$ and $\tau$  channels.} \label{Fig:difff}
 \end{figure}

  In addition, it is also important to plot the differential decay widths, for which the lepton can be set to be electron, muon and tau, such that the impact of the lepton mass can be assessed. The differential decay widths of  $cc$ and $bb$ sector  as  functions of \( q^2 \) for all possible lepton channels can be seen in figures \ref{Fig:diff}  and \ref{Fig:difff}, respectively.
  
\section{Conclusion}~\label{Sec5}

In conclusion,  we  investigated the semileptonic decays of the doubly charmed (bottom) baryons into the single heavy baryons using the QCD sum rule approach in all possible lepton channels.  We considered the interpolating currents of the involved initial and final baryons with their most possible general forms.  Our calculations accounted for nonperturbative operators with mass dimensions up to five, yielding transition form factors characterized through the vector and axial vector transition currents. Following the determination of working intervals of the auxiliary parameters, we derived the fit functions for six relevant form factors with respect to \( q^2 \) within the allowed physical ranges at all considered decay channels. These findings enabled us to estimate decay widths and branching ratios for all possible lepton channels.  It should be noted that in  the computations we  utilized the newly  obtained residue and mass  values for doubly heavy baryons \cite{ShekariTousi:2024mso}. We estimated  the decay rates and branching fractions for each lepton channel separately, achieving results with minimized uncertainties through careful auxiliary parameter selection. 

The identification of the doubly charmed baryon \( \Xi_{cc} \) has spurred extensive theoretical investigations into hadron spectroscopy and the decay mechanisms associated with baryons containing two heavy quarks. As experiments like LHCb continue to search for these particles, our analysis emphasizes the need for a deeper understanding of their properties.  Despite the intensive theoretical expectations about the  doubly heavy baryons, only \( \Xi^{++}_{cc} \) has been experimentally confirmed to date. Given advancements from experiments such as LHCb at CERN, we anticipate the soon-to-come discovery of other members of the baryons containing two heavy quarks. Our findings will support experimental teams in this pursuit. The orders of magnitude obtained for  branching fractions for the examined weak decays indicate that these semileptonic processes can be achieved and  investigated in experiments such as LHCb in near future.  Future experimental results related to these observables, compared with our theoretical predictions, will provide crucial information about the internal composition and structure of doubly heavy baryons. This will also allow for accurate evaluations of essential characteristics of these interesting baryons like their  mass and lifetime.

\section*{ACKNOWLEDGMENTS}

K. Azizi thanks Iran national science foundation (INSF) for the partial financial support supplied under the
elites Grant No. 4037888. He is also thankful to the CERN-Theory department  for their support and  hospitality.

\section*{APPENDIX: PERTURBATIVE AND NONPERTURBATIVE CONTRIBUTIONS }

The explicit expressions for the perturbative and nonperturbative components of the  spectral density for the semileptonic transition  $ \Omega_{bb}^{-}\rightarrow \Xi^0_{b} {\ell} \bar{\nu_{\ell}}$, as an example,  for the structure $\gamma_{\mu} \gamma_5$ are shown as:
\begin{eqnarray} \label{RhoPert}
&&\rho^{Pert.}_{\gamma_{\mu} \gamma_5}(s,s',q^2)=\int_{0}^{1}du \int_{0}^{1-u}dv \int_{0}^{1-u-v}dz~\frac{\sqrt{3}  }{256 A H^3 J^6 \pi^4} \notag\\
&& \bigg\{-D_1^2 J^4 \bigg[B (11 + 7 \beta) H (H - 2 u) v + 
z \bigg(-\big((11 + 7 \beta) C H (1 + 2 u)\big) - 
2 u z \big(18 u + 7 (-1 + \beta) z\big) + 
H \big(v (-40 - 18 u \notag\\
&&+ 7 \beta (-2 + v) + 29 v) + \big(-22 + 3 u + 
29 v + 7 \beta (-2 - 3 u + v)\big) z + (11 + 
7 \beta) z^2\big)\bigg)\bigg] - 
2 A H^2 v^2 z \bigg[H J^2 m_b^2 \big(A \beta s' \notag\\
&&+ 4 (-q^2 + s) u - 
s' (5 H + u + 5 z)\big) + 
s' z \bigg\{-\big[(-H + u - z) \big((5 + \beta) H s' + 2 (2 + \beta) s' u + 
q^2 (u - \beta u) + (5 + \beta) s' z\big)\notag\\
&& \big(B H v + 
z (C H + 2 H v + H z + 2 u z)\big)\big] - 
s \bigg[B H \big(3 (1 + \beta) H^2 + 2 (2 + \beta) H u - 
6 (2 + \beta) u^2\big) v + 
z \bigg(H \big(-3 (1 + \beta) \notag\\
&&+ (7 + 5 \beta) u + 4 (2 + \beta) u^2 - 
6 (2 + \beta) u^3 + \big[15 - 7 u (4 + u) + 
\beta \big(15 + (-20 + u) u\big)\big] v + 
3 (-7 - 7 \beta + 7 u + 5 \beta u) v^2 \notag\\
&&+ 
9 (1 + \beta) v^3\big) + \big(6 (-1 + \beta) u^3 + (7 + 
5 \beta) H u (-2 + 3 v) + 
3 (1 + \beta) H^2 (-3 + 4 v)\big) z + \big(9 (1 + 
\beta) H^2 + (7 + 5 \beta) H u \notag\\
&&+ 
4 (2 + \beta) u^2\big) z^2 + 3 (1 + \beta) H z^3\bigg)\bigg]\bigg\}\bigg] - 
2 D_1 H J^2 v \bigg\{2 A (-4 + \beta) H J^2 m_b^2 + 
u z \bigg[q^2 \bigg\{B H v \big(-9 + 13 u + \beta (3 + 13 u - 3 v) \notag\\
&&+ 9 v\big) + 
z \bigg(C H (3 (-3 + \beta) + 13 (1 + \beta) u) + 
H v \big(4 (-9 + 3 \beta + 11 u + 5 \beta u) - 
9 (-3 + \beta) v\big) + (-6 (-3 + \beta) - 
8 (5 + \beta) u \notag\\
&&+ 2 (13 + 15 \beta) u^2 + 
15 (-3 + \beta) v + 8 (5 + \beta) u v - 
9 (-3 + \beta) v^2) z + \big(9 (H + 2 u) + 
\beta (3 - 2 u - 3 v)\big) z^2\bigg)\bigg\} + 
s \bigg(B H v (5 + \beta \notag\\
&&+ 2 u + 2 \beta u - (5 + \beta) v) + 
z \big[C H \big(5 + \beta + 2 (1 + \beta) u\big) + 
H v \big(24 - 4 (4 + 3 \beta) u + (-19 + \beta) v\big) + \bigg(-10 - 
2 \beta + 21 u \notag\\
&&- 3 \beta u - 16 u^2 - 
28 \beta u^2 + \big(29 + \beta + 
3 (-7 + \beta) u\big) v + (-19 + \beta) v^2\bigg) z + \big(5 + 
\beta - 18 u + 2 \beta u - (5 + \beta) v\big) z^2\big]\bigg)\bigg] + 
2 s' z \notag\\
&&\bigg[2 B H v \big(5 + \beta + 5 u + \beta u - 11 u^2 - 
6 \beta u^2 - (5 + \beta) (2 + u) v + (5 + \beta) v^2\big) + 
z \bigg(2 H^3 \big(-5 - \beta + 3 (7 + \beta) v \big) \notag\\
&&+ 
H^2 \big(-3 (-1 + \beta) u v - 6 (5 + \beta) z + 
8 (8 + \beta) v z + 6 \beta z^2\big) + 
H \bigg(-u^2 \big(-32 + 22 u + 63 v + \beta (-14 + 12 u + 29 v)\big) \notag\\
&&- 
u \big(21 + 52 u + \beta (3 + 20 u) - 24 v\big) z + 
6 (-5 + 7 u + 7 v) z^2\bigg) + 
z \big[-5 (9 + 5 \beta) u^3 - 2 (10 + 3 \beta) u^2 z + 
2 (5 + \beta) (-1 + v) z^2 \notag\\
&&+ 
3 u z \big(2 \beta (-1 + v) + (7 + \beta) z\big)\big]\bigg)\bigg]\bigg\} + 
\beta \bigg[-D_1^2 J^4 \bigg(B (7 + 11 \beta) H (H - 2 u) v + 
H \bigg(-\big((7 + 11 \beta) C (1 + 2 u)\big) + 7 (-2 + v) v \notag\\
&&+ 
\beta v (-40 - 18 u + 29 v)\bigg) z + \big(-36 \beta u^2 + 
7 H (-2 - 3 u + v) + 
\beta H (-22 + 3 u + 29 v)\big) z^2 + \big((7 + 11 \beta) H + 
14 (-1 + \beta) u\big) \notag\\
&&z^3\bigg) + 
2 D_1 H J^2 v \bigg[2 A (-4 + \beta) H J^2 m_b^2 - 
z \bigg(4 B H s' v (1 + 5 \beta + u + 5 \beta u - 6 u^2 - 
11 \beta u^2 - (1 + 5 \beta) (2 + u) v + (1 + 
5 \beta) v^2) \notag\\
&&+ 
q^2 u \bigg[B H v \big(3 + 13 u - 3 v + \beta (-9 + 13 u + 9 v)\big) + 
z \bigg(C H \big(3 + 13 u + \beta (-9 + 13 u)\big) + 
H v \big(12 + 20 u - 9 v + 
\beta (-36 + 44 u \notag\\
&&+ 27 v)\big) + \big(-6 + 8 H u + 
30 u^2 + 26 \beta u^2 + 15 v - 9 v^2 + 
\beta H (-18 + 40 u + 27 v)\big) z + \big(3 - 2 u + 
9 \beta (H + 2 u) - 3 v\big) z^2\bigg)\bigg] \notag\\
&&+ 
s u \bigg[B H \big(-H + 2 u + \beta (5 + 2 u - 5 v)\big) v + 
z \big\{C H \big(1 + 5 \beta + 2 (1 + \beta) u\big) + 
H v \big(-12 u + \beta (24 - 16 u - 19 v) + v\big) + \bigg(-2\notag\\
&& + 
3 H u - 28 u^2 + v + v^2 - 
\beta \big(16 u^2 + H (-10 + 21 u + 19 v)\big)\bigg) z - \big(H - 
2 u + \beta (-5 + 18 u + 5 v)\big) z^2\big\}\bigg] + 
2 s' z \big[2 H^3 \big(-1 + 3 v \notag\\
&&+ \beta (-5 + 21 v)\big) + 
z \big(-5 (5 + 9 \beta) u^3 - 
2 (3 + 10 \beta) u^2 z + (-2 + 3 u + 21 \beta u + 
2 v) z^2\big) + 
H \big[-u^2 \big(-14 + 12 u + 29 v + \beta (-32 \notag\\
&&\notag\\
&&+ 22 u + 63 v)\big) -
u \big(3 + 20 u + \beta (21 + 52 u - 24 v)\big) z + 
6 \big(u + \beta (-5 + 7 u + 7 v)\big) z^2 + 10 \beta z^3\big] + 
H^2 \big(3 (-1 + \beta) u v \notag\\
&&\notag\\
&&
+ 
2 z (-3 - 15 \beta + 4 v + 32 \beta v + 3 z)\big)\big]\bigg)\bigg]+ 
2 A H^2 v^2 z \bigg(H J^2 m_b^2 \big(A \beta s' + 4 (-q^2 + s) u - 
s' (5 H + u + 5 z)\big) + 
s' z \big\{(-H + u \notag\\
&&- z) \big[(-1 + \beta) q^2 u + 
s' \big(H + 2 u + 4 \beta u + 5 \beta (F + v) + z\big)\big] \big(B H v + 
z (C H + 2 H v + H z + 2 u z)\big) + 
s \big[B H (3 (1 + \beta) H^2 \notag\\
&&- 6 (1 + 2 \beta) u^2 + 
2 H \big(u + 2 \beta u)\big) v + 
z \big(C H \big[3 - 2 u (1 + 3 u) + 
\beta \big(3 - 4 u (1 + 3 u) + 21 v^2\big)\big] + 
3 (1 + \beta) H^2 z (-3 + 4 v + 3 z) \notag\\
&&
+2 u^2 z (3 u - 3 \beta u + 2 z + 4 \beta z) + 
H \{3 (-7 + 5 u) v^2 + 
9 (1 + \beta) v^3 + (5 + 7 \beta) u (-2 + z) z + 
3 (1 + \beta) z^3 + 
v [15 + \beta [15 \notag\\
&&- 7 u (4 + u - 3 z)] + 
u (-20 + u + 15 z)]\}\big)\big]\big\}\bigg)\bigg] \bigg\} \Theta[D_1(s,s^{\prime},q^2)], 
\end{eqnarray}

\begin{eqnarray} \label{Rho3}
&&\rho^3_ {\gamma_{\mu} \gamma_5}(s,s',q^2)=
\int_{0}^{1}du \int_{0}^{1-u}dv ~\frac{ 1 }{32 \sqrt{3}  C^3 \pi^2} \notag\\
&&\bigg[-\bigg(\big(11 + \beta (14 + 11 \beta)\big) m_b \langle \bar{s} s\rangle \Theta[L(s,s^{\prime},q^2)]\bigg) + 
2 (1 + \beta - 2 \beta^2) C^3 m_b \langle \bar{u} u\rangle \Theta[L'(s,s^{\prime},q^2)]\bigg],
\end{eqnarray}

\begin{eqnarray}\label{Rho4}
&&\rho^4_{\gamma_{\mu} \gamma_5}(s,s',q^2)=\int_{0}^{1}du \int_{0}^{1-u}dv \int_{0}^{1-u-v}dz\frac{1}{1024 H^5 J^4 \pi^2} \langle0|\frac{1}{\pi}\alpha_s G^2|0\rangle \bigg\{2 H^5 \bigg[2 (1 + \beta^2) F^2 u z - 
2 \big(19 + \beta (12 + 11 \beta)\big) \notag\\
&&u^3 (v + z) + (1 + \beta)^2 v^2 (H +
z) (H + 9 z) + 
u v^2 (2 (1 + \beta^2) (-2 + v) + (35 + 
3 \beta (14 + 9 \beta)) z) + 
F \bigg(-4 \big(9 + \beta (6 + 5 \beta)\big)  \notag\\
&&u^2 z+ (1 + \beta)^2 v z (H + 
z) + 2 u v \big(-1 - 
\beta^2 + [13 + 3 \beta (4 + 3 \beta)] z\big)\bigg) - 
2 u^2 v \bigg(-18 + 19 v + 27 z + 
\beta \big(2 (-6 + 7 v + 7 z)\notag\\
&& + 
\beta (-10 + 11 v + 15 z)\big)\bigg)\bigg]  + 
z \bigg[H^4 v^2 \bigg(4 u (-1 + 2 u - 5 v) + 2 H v + 
2 \beta^2 [2 u (-1 + 2 u - 5 v) + H v] \notag\\
&&- 
\beta \big(14 + 3 (-13 + u) u - 26 v + 55 u v + 12 v^2 + 
2 H (-7 + 6 v)\big)\bigg) + 
H^4 v \bigg(-6 \big(1 + \beta (6 + \beta)\big) + 2 (1 + \beta (45 + \beta)) u \notag\\
&&+ 
2 \big(8 + \beta (-3 + 8 \beta)\big) u^2 - \bigg(2 + 50 u + 
\beta \big(-92 + 201 u + \beta (2 + 50 u)\big)\bigg) v+ 
16 \big(1 + (-3 + \beta) \beta \big) v^2\bigg) z  \notag\\
&&+ 
H^2 \bigg[8 u^2 \big(1 + 2 (-2 + v) v\big) - 4 u (1 + v - 17 v^2 + 9 v^3) + 
4 H v \big(-1 + v (-3 + 2 v)\big) + 
4 \beta^2 \bigg(u^2 \big(2 + 4 (-2 + v) v\big)  \notag\\
&&- 
u (1 + v - 17 v^2 + 9 v^3) + H v \big(-1 + v (-3 + 2 v)\big)\bigg) + 
\beta \bigg(u \big(39 - 3 u \big(1 + 2 (-2 + v) v\big) + 
v (-369 + 574 v - 228 v^2)\big)  \notag\\
&&- 
4 H \big[7 + v \big(49 + v (-71 + 31 v)\big)\big]\bigg)\bigg] z^2 + 
H^2 \bigg(\beta (96 + 6 u^2 + 8 v \big(-44 + (49 - 18 v) v\big) + 
u (-157 + 346 v - 78 v^2)) \notag\\
&& + 
4 \bigg(-4 u^2 + v \big(3 + (-6 + v) v\big) + 2 u \big(1 + v (4 + v)\big)\bigg) + 
4 \beta^2 \big[-4 u^2 + v \big(3 + (-6 + v) v\big) + 
2 u \big(1 + v (4 + v)\big)\big]\bigg) z^3  \notag\\
&&- 
2 H \big[-2 u + 
\beta \big(-27 + 27 H + (59 - 2 \beta) u\big) + \bigg(5 H + 
\beta \big(66 + (-39 + 5 \beta) H\big) + 4 u + 
4 \beta (10 + \beta) u  \notag\\
&&+ (8 + 
\beta \big(-3 + 8 \beta)\big) u^2\bigg) v + \bigg(11 H - 26 u + 
\beta \big(-56 + (17 + 11 \beta) H - (115 + 26 \beta) u\big)\bigg) v^2 + 
17 \beta v^3\big] z^4  \notag\\
&&+ 
2 H \bigg(-8 (u - 2 v) (u - v) + 8 v + 
8 \beta^2 \big(-\big[(u - 2 v) (u - v)\big] + v\big) + 
\beta \big(-8 + 3 u^2 + 12 H v + 
u (-79 + 174 v)\big)\bigg) z^5  \notag\\
&&+ \bigg(-8 (u^2 + 3 H v - 3 u v) - 
8 \beta^2 (u^2 + 3 H v - 3 u v) + 
\beta \big(60 H^2 + 3 (-79 + u) u + 253 u v\big)\bigg) z^6 + \bigg(\beta (-48 + 
79 u)  \notag\\
&&- 8 \big(1 + (-5 + \beta) \beta\big) v\bigg) z^7 + 
12 \beta z^8\bigg]\bigg\} \Theta[D_2(s,s^{\prime},q^2)],
\end{eqnarray}

\begin{equation}\label{Rho5}
	\rho^5_{\gamma_{\mu} \gamma_5}(s,s',q^2)=0,
\end{equation}
%%%%%
where,
\begin{eqnarray}\label{L}
&&D_1(s,s^{\prime},q^2)=\frac{-H}{J^2} \bigg(-q^2 u (v + z) A + 
	m_b^2 (u + v) (v^2 + v F + F z) + 
	v z (s u - s' A)\bigg),
\\
&&D_2(s,s^{\prime},q^2)=\frac{-H }{J^2} \bigg(-q^2 u (v + z) A + 
	m_b^2 (u + v) J + 
	v z (s u - s' A)\bigg),
\\	
&&L=\frac{1}{H^2} \bigg(-s C u + (-q^2 + s + s') C u + m_b^2 C (u + v) - 
s' (u^2 + u H + H v)\bigg),
\\
&&L'=-s C u + (-q^2 + s + s') u B - m_b^2 (u + v) + 
s' (u - u^2 + v - 2 u v - v^2),
\end{eqnarray}
and $\Theta[...]$ represents  the unit step function.
We have used
\begin{eqnarray}\label{dL}
&&J=v^2 + v F + F z,\notag
\\
&&A=-1 + u + v + z,\notag
\\
&&B=-1 + u + v,\notag
\\
&&C=-1 + u,\notag
\\
&&F=-1 + z,\notag
\\
&&H=-1 + v.
\end{eqnarray}
%

%%%%%%%%%%%%%%%%%%%%%%%%%%%%%%%%%%%%%%%%%%%%%%%%%%%%%%%%%%%
%%%%%%%%%%%%%%%%%%%%%%%%%%%%%%%%%%%%%%%%%%%%%%%%%%%%%%%%%%%%%%%%%%%%%%%%%%%%%%%%%%%%%%%%%%%%%%%%%%%%%%%%%%%%

\end{document}